\documentclass{ecai}
\usepackage{times}
\usepackage{graphicx}
\usepackage{tabularx}
\usepackage{latexsym}
\usepackage{braket}
\usepackage{bm}
\usepackage{hyperref}
\usepackage{pgfplots}
\usepackage{booktabs}
\usepackage{multirow}
\usepackage[ruled,vlined]{algorithm2e}
\usepackage{amsmath,amsthm,amssymb} 
\usepackage[normalem]{ulem}

\ecaisubmission   

\begin{document}

\title{Finding Quantum Critical Points with Neural-Network Quantum States}

\author{
Remmy Zen\institute{School of Computing, National University of Singapore,
Singapore, emails: remmy@u.nus.edu, steph@nus.edu.sg} \and 
Long My\institute{Centre for Quantum Technologies, National University of Singapore, Singapore, email: e0008987@u.nus.edu} \and 
Ryan Tan\institute{Engineering Product Development Pillar, Singapore University of Technology and Design, Singapore, emails: ryan\_tan@mymail.sutd.edu.sg, dario\_poletti@sutd.edu.sg} \and 
Fr\'ed\'eric H\'ebert\institute{Universit\'e C\^ote d’Azur, CNRS, INPHYNI, Nice, France, emails: \{frederic.hebert, mario.gattobigio\}@inphyni.cnrs.fr} \and 
Mario Gattobigio$^{4}$ \\ \and 
Christian Miniatura\institute{MajuLab, CNRS-UCA-SU-NUS-NTU International Joint Research Unit, Singapore, email: cqtmc@nus.edu.sg}\hspace{4pt}$^{,2,}$\institute{Department of Physics, National University of Singapore, Singapore}\hspace{4pt}$^{,}$\institute{School of Physical and Mathematical Sciences, Nanyang Technological University, Singapore}\hspace{4pt}$^{,}$\institute{Yale-NUS  College,  16  College  Avenue  West,  Singapore  138527,  Singapore}\hspace{4pt}$^{,4}$
\and 
Dario Poletti\institute{Science and Mathematics Cluster, Singapore University of Technology and Design, Singapore}\hspace{4pt}$^{,3,5}$ \and 
St\'ephane Bressan$^{1}$}

\maketitle
\bibliographystyle{ecai}

\begin{abstract}
Finding the precise location of quantum critical points is of particular importance to characterise quantum many-body systems at zero temperature. However, quantum many-body systems are notoriously hard to study because the dimension of their Hilbert space increases exponentially with their size. Recently, machine learning tools known as neural-network quantum states have been shown to effectively and efficiently simulate quantum many-body systems.   
We present an approach to finding the quantum critical points of the quantum Ising model using neural-network quantum states, analytically constructed innate restricted Boltzmann machines, transfer learning and unsupervised learning. We validate the approach and evaluate its efficiency and effectiveness in comparison with other traditional approaches. 
\end{abstract}


\section{INTRODUCTION}
Matthias Vojta, in~\cite{vojta2003quantum}, highlights that ``[...] the presence of [...] quantum critical points holds the key to so-far unsolved puzzles in many condensed matter systems''.
Quantum critical points mark the transition between different phases of  quantum many-body systems~\cite{thouless2014quantum} at zero temperature. Finding their precise location is of particular importance to characterise the physical properties of quantum many-body systems~\cite{sachdev2007quantum,sondhi1997continuous}. However, these systems are notoriously hard to study because their associated quantum wave functions live in a huge Hilbert space which is the tensor product of the individual Hilbert spaces associated to each constituent of the system. As such, its dimension increases exponentially with the number of constituents. This entails computational complexity issues and calls for deterministic and stochastic approximation algorithms. 

Recently, Carleo and Troyer~\cite{carleo2017solving} showed that a machine learning tool, which they called \emph{neural-network quantum states}, can effectively and efficiently simulate quantum many-body systems in different quantum phases and for different parameters of the system. Their approach can be seen as an unsupervised neural network implementation of a variational quantum Monte Carlo method. The authors used a restricted Boltzmann machine to calculate the ground state energy and the time evolution of quantum many-body systems such as the Ising and Heisenberg models. This work triggered a wave of interest in the design of neural network approaches to the study of quantum many-body systems~\cite{choo2018symmetries,choo2019fermionic,choo2019two,jonsson2018neural,mcbrian2019ground,melko2019restricted, westerhout2019neural, zen2019transfer}.

We present here an approach to finding the quantum critical points of the quantum Ising model using innate restricted Boltzmann machines, transfer learning and unsupervised learning for neural-network quantum states. 

We first propose to analytically construct restricted Boltzmann machine neural-network quantum states  for quantum states deeply in each phase of the system. We refer to such restricted Boltzmann machine neural-network quantum states as \emph{innate} as they have innate knowledge, i.e. built-in knowledge rather than knowledge acquired by training, of the system they represent.

We then devise a transfer learning protocol across parameters of the system to improve both the efficiency and the effectiveness of the approach. We finally combine the transfer learning protocol across system parameters with a transfer learning protocol to larger sizes~\cite{zen2019transfer} to find the quantum critical points in the limit of infinite size.

We evaluate the efficiency and effectiveness of the approach for one-, two- and three-dimensional Ising models in comparison with other traditional approaches such as exact diagonalization method~\cite{weisse2008exact},  a numerical approximation method called tensor network method~\cite{orus2014practical,schollwock2011} and a stochastic method called quantum Monte Carlo method~\cite{blote2002cluster,braiorrorrs2016phase,gubernatis2016}.

The rest of the paper is structured as follows. Section~\ref{sec:relatedwork} summarises the related work. Section~\ref{sec:nqs} presents the necessary notions of quantum many-body physics, the Ising model, restricted Boltzmann machine and restricted Boltzmann machine neural-network quantum states. Section~\ref{sec:transferlearning} presents the general approach and the algorithm for finding quantum critical points, the transfer learning protocols and the analytical construction of an initial restricted Boltzmann machine neural-network quantum states. Section~\ref{sec:performanceeval} reports the result of the comparative performance evaluation of our approach. We conclude and highlight possible future works in Section~\ref{sec:conclusion}.


\section{RELATED WORK} \label{sec:relatedwork}

\subsection{Quantum many-body physics} \label{subsec:related_manybody}
Quantum many-body physics~\cite{thouless2014quantum} is a branch of quantum physics that studies  systems with large numbers of interacting particles, or bodies. Some well-known quantum many-body physics models are Ising, Heisenberg and Hubbard models and their variants~\cite{affleck2004rigorous,elliott1961phenomenological,gersch1963quantum}. We focus on the quantum Ising model, which has been studied extensively in the literature~\cite{bravyi2017complexity,suzuki2012quantum,white1983quantum} as it, albeit a simple model, displays most of the qualitative features present in complex models.

More specifically we focus on  finding the ground state of a system in the Ising model and the quantum critical points where the nature and the properties of the ground state change qualitatively.

Several methods have been developed to find the ground state. The most straightforward method is to diagonalize the Hamiltonian matrix that represents the problem, its eigenvalues giving the possible energies and its eigenvectors representing the corresponding states~\cite{weisse2008exact}. 

Even though iterative methods~\cite{lanczos1950iteration,lehoucq1998arpack} have been devised to improve the efficiency of the diagonalization method, it still does not scale well as the size of the system increases. Instead, deterministic and stochastic approximation methods have been proposed and used. Tensor network methods~\cite{schollwock2011} are deterministic approximation methods using variational techniques and combining the exact diagonalization with the iterative generation of an effective, low dimensional and local Hamiltonian~\cite{white1992density}. Quantum Monte Carlo methods~\cite{gubernatis2016} are, instead, stochastic.

This work belongs to the general field of quantum machine learning that addresses machine learning problems with quantum computing as well as quantum physics problems with machine learning~\cite{biamonte2017quantum,carleo2019review,sarma2019review}. One application is the evaluation of the properties of a quantum system at very cold temperatures \cite{carleo2017solving}. Some other applications in this domain include quantum state tomography which reconstructs quantum states from  measurements~\cite{torlai2018neural}, the estimation~\cite{jarzyna2015true} and control~\cite{chen2014fidelity} of the parameters of quantum systems, and the design of better experiments~\cite{august2017using}. Machine learning algorithms, specifically neural network, have also been used to classify phases or to detect phase transitions in a supervised~\cite{broecker2017machine,carrasquilla2017machine,rem2019identifying,van2017learning} and unsupervised~\cite{hu2017discovering,wang2016discovering,wetzel2017unsupervised} manner. 

\subsection{Neural-network quantum states}

Recently, Carleo and Troyer proposed to use restricted Boltzmann machines to simulate quantum many-body systems and introduced neural-network quantum states~\cite{carleo2017solving}. Their scheme falls into the family of variational quantum Monte Carlo methods. They tested their approach on the paradigmatic Ising and Heisenberg models and demonstrated that this new method is capable of finding the lowest-energy state and of reproducing the time evolution of these interacting systems. The neural-network quantum states method has been further explored by studying quantum entanglement properties~\cite{deng2017quantum}, its connection to other methods~\cite{chen2018equivalence,glasser2018neural} and its representation power~\cite{collura2019descriptive,gao2017efficient,jia2019efficient,lu2019efficient}. It has also been used to find excited states~\cite{choo2018symmetries}, to study different quantum models~\cite{choo2019fermionic,choo2019two,mcbrian2019ground} and to aid the simulation of quantum computing~\cite{jonsson2018neural}.

Several works have tried to analytically or algorithmically construct a representation of a quantum many-body system with a restricted Boltzmann machine. Most of these works focus on the architecture and topology of the network.
Restricted Boltzmann machines have been constructed for the Majumdar-Ghosh and AKLT models~\cite{glasser2018neural}, for the CZX model~\cite{lu2019efficient} and for the Heisenberg and Hubbard models (in this case combined with the pair-product method)~\cite{nomura2017restricted}. In the field of quantum error correction, restricted Boltzmann machines have been proposed for the stabilizer code~\cite{jia2019efficient} and for the toric code~\cite{gao2017efficient}. The authors of~\cite{carleo2018constructing} algorithmically and deterministically constructed a deep Boltzmann machine from the system parameters. The authors of~\cite{chen2018equivalence} algorithmically constructed a mapping between restricted Boltzmann machines and tensor networks. 

\subsection{Restricted Boltzmann machine}

The original architecture of neural-network quantum states, which we adopt, leverages the unsupervised training  and  generative capabilities of restricted Boltzmann machines. Restricted Boltzmann machines are generative energy-based probabilistic graphical models. They were initially invented under the name Harmonium in 1986~\cite{smolensky1986information}.

The training of restricted Boltzmann machines can be supervised or unsupervised. In the supervised case, they are usually used as feature extractors. However, they can also be used for classification and regression tasks as in~\cite{larochelle2008classification,lu2016deep,midhun2014deep,nguyen2017supervised,tomczak2015classification}. In the unsupervised case, they have been used in a variety of domains such as face recognition~\cite{teh2001rate}, dimensionality reduction~\cite{hinton2006reducing}, unsupervised feature learning~\cite{coates2011analysis} and image denoising~\cite{tang2012robust}, topic modelling~\cite{hinton2009replicated,xie2015diversifying}, acoustic modelling~\cite{dahl2010phone,jaitly2011learning}, collaborative filtering~\cite{salakhutdinov2007restricted}, anomaly detection~\cite{fiore2013network}, fault detection~\cite{liao2016enhanced} and credit scoring~\cite{tomczak2015classification}.

One of their most interesting characteristics is that they can, not only be used as discriminative models, but also as generative models~\cite{boulanger2012modeling,kivinen2012multiple,le2011learning,schmah2009generative,sutskever2009recurrent,taylor2007modeling,wu2013conditional}. They have been applied to the generation of images~\cite{le2011learning,kivinen2012multiple}, videos~\cite{sutskever2009recurrent}, music~\cite{boulanger2012modeling}, speeches~\cite{wu2013conditional} and human motions~\cite{taylor2007modeling}. The authors of~\cite{schmah2009generative} use the combination of generative and discriminative models of restricted Boltzmann machines in the medical domain to classify and generate fMRI images.

\subsection{Transfer learning}

We devise two transfer learning protocols that improve effectiveness, efficiency and scalability of restricted Boltzmann machine neural-network quantum states.

Gale Martin, in~\cite{martin1988effects}, was the first to evaluate the opportunity of directly copying neural network weights trained on a particular task to another neural network with a different task in order to improve efficiency. His approach was soon further improved by~\cite{pratt1993discriminability}. The notion was later formalised in~\cite{thrun1998learning} and in~\cite{baxter1998theoretical} under the name \emph{transfer learning}. 

Transfer learning has been applied to all kinds of unsupervised, supervised and reinforcement learning tasks as reported in several surveys~\cite{pan2010survey,weiss2016survey}. Transfer learning has been applied to restricted Boltzmann machines for numerous tasks such as reinforcement learning~\cite{ammar2013automatically} and classification~\cite{wei2011heterogeneous,zhang2011deep}. The authors of~\cite{yosinski2014transferable} applied transfer learning to neural networks and they observed that it improves both efficiency and effectiveness.


\section{NEURAL-NETWORK QUANTUM STATES}  \label{sec:nqs}
\subsection{Quantum many-body systems}\label{subsec:manybody}
A quantum many-body system~\cite{thouless2014quantum} consists of a large number of interacting particles, or bodies, evolving in a discrete or continuous $D$-dimensional space.
A particle is, in general, characterised by its external degrees of freedom, such as its position  momentum, and its internal degrees of freedom, such as its magnetic moment, also referred to as spin. In the following, we will concentrate on the spin degree of freedom and consider identical particles pinned at the nodes of a $D$-dimensional lattice ($D=1,2,3$). The size of the system is then given by the number $N$ of particles, the number of possible spin states per particle being $n_s$.

A quantum many-body model defines how particles interact with each other or with external fields. Several prototypical models, such as the Ising and Heisenberg models and their variants~\cite{elliott1961phenomenological,gersch1963quantum}, describe the pairwise interactions of the spins of particles in addition to the interaction with external fields. The physical properties of each model depend on the respective magnitude of all these interactions which enter as parameters in the model.

Specifying the value of the spin for each particle gives a configuration of the system.The number of configurations $n_s^N$ is, therefore, exponential in the number of particles. We will specifically consider one-half spin in the rest of the paper, meaning that each particle can only have $n_s=2$ internal states.

In quantum physics, the possible physical states of a given system are described by state vectors $|\Psi\rangle$, called wave functions, living in the so-called state space. Formally, this state space is a complex separable Hilbert space and state vectors are simply linear combination of all the basis state vectors, denoted by $|x\rangle$, associated to each possible configuration $x$:

\begin{equation}
\label{eq:expansion}
 |\Psi\rangle = \sum_{x} \Psi(x) |x\rangle
\end{equation}

As easily seen from Equation \ref{eq:expansion}, the dimension of the Hilbert space is given by the number of possible distinct configurations. For the interacting spin systems we consider, the dimension of the Hilbert space is then $2^N$. 
Each complex coefficient $\Psi(x)$ in Equation \eqref{eq:expansion} is called a probability amplitude. Defining the normalisation constant $Z_{\Psi}=\sum_x |\Psi(x)|^{2}$, $|\Psi(x)|^{2} /{Z_{\Psi}}$ gives the probability of the configuration $x$ in the state $|\Psi\rangle$. The collection of all these probabilities defines the multinomial probability distribution of all possible configurations $x$ of the system.

For a given grid, number of particles and external fields, the dynamics of a system is fully described by its Hamiltonian. The Hamiltonian is a Hermitian matrix of size $n_s^N \times n_s^N$ that describes how the system evolves. Furthermore, the eigenvalues of the Hamiltonian are the possible energies of the system and the corresponding eigenvectors are the only possible states in which the system can be individually found after a measurement of its energy has been performed.

The energy functional $E[\Psi] = \langle\Psi | H | \Psi\rangle / Z_\Psi$ of a state with wave function $|\Psi\rangle$ is given in Equation~\ref{eq:energyfunction} where $E_{loc}$ is the local energy function of a given configuration $x$, as defined in Equation~\ref{eq:eloc}, with $H_{x,x'}=\langle x | H | x' \rangle$ the entry of the Hamiltonian matrix for the configurations $x$ and $x'$: 

\begin{equation}
\label{eq:energyfunction}
 E[\Psi] = \sum_{x} \frac{|\Psi(x)|^{2}}{Z_{\Psi}} E_{loc}(x) 
\end{equation}
 
 \begin{equation}
\label{eq:eloc}
 E_{loc}(x) = \sum_{x'} H_{x,x'} \frac{\Psi(x')}{\Psi(x)}
\end{equation}

Formally, the energy functional is the expected value of the local energy. Do note that the local energy $E_{loc}$ of any configuration $x$ gives the average energy value of the corresponding state $|x\rangle$. Based on the variational principle in quantum mechanics, the energy functional of a given state $|\Psi\rangle$ is always larger than or equal to the lowest possible energy of the system, i.e. to the lowest eigenvalue of the Hamiltonian. It reaches this minimal value when $|\Psi\rangle$ is precisely the corresponding eigenvector called the ground state of the system.

Being the most relevant state at low enough temperatures, the ground state has important physical implications as it can have emerging properties which could not be trivially predicted from the interactions of the particles.

A phase is a region in the space of the parameters of a model in which systems have similar physical properties. In the thermodynamic limit, each possible phase is characterised by so-called order parameters that achieve different values in each phase region. Finding the order parameters that characterise the phases of a system is an open research area which can benefit from neural networks too~\cite{carleo2019review, sarma2019review}.  

A phase transition occurs when the system crosses the boundary between two phases and the order parameters change values. When this happens, the nature and the properties of the ground state change qualitatively. The transition happens when the parameters of a model are varied.  In quantum systems, in the limit of infinite system size, the transition is typically described by an abrupt change in the observable physical properties or their derivatives. In particular, the term ``quantum phase transition'' is used for phase transitions in the ground state alone (i.e. for a system at zero temperature). The parameters of a model that correspond to this abrupt change define the quantum critical points. For finite-size systems, the transition is not abrupt but smooth. Mathematically, this means that, for a given size of the system, we need to find the inflection point of the order parameter as a function of the parameters of the system. Since it is not possible to empirically determine the parameters that yield the quantum critical point of an infinite system, it will be necessary to extrapolate its limit value from a series of values measured or simulated from systems of increasing sizes. In the remainder of the paper, when we mention a critical point, we refer to the quantum critical point. 

\subsection{Ising Model} \label{subsec:ising}

The Ising model describes particles pinned on the sites of a lattice carrying a binary discrete spin. Each spin is in one of two states: up (represented by $+1$) or down (represented by $-1$). We have two states per particles and, therefore, the number of configurations equals to $2^N$, for $N$ particles. A configuration $x$ is given by the value of the spin on each site:
$x=(x_1,x_2, \cdots, x_N)$ where $x_i = \pm 1$.

Each particle interacts with its nearest neighbours and with an external magnetic field along the $x$-axis. We consider a homogeneous Ising model where the parameters are translationally invariant. The parameters of the model that characterise the interaction among the particles and the external field are denoted with $J$ and $h$, respectively. 

Equation~\ref{eq:hamiltonian-tfi} gives the $2^N\times 2^N$ Hamiltonian matrix of the Ising model where $neigh(\cdot)$ is a function that returns the nearest neighbouring sites and the $\sigma^\alpha_i$ are the Pauli matrices where $\alpha=x,y,z$ and $i$ indicates the position of the spin it acts upon.  
Only the relative strength between $J$ and $h$ matters. For instance, a realisation of the Ising model with $h = 1$ and $J = 1$ has the same static properties as a realisation  with $h = 2$ and $J = 2$ except that the energy is doubled in the latter. Therefore, we refer to $J/|h|$ as the parameter of the system in the Ising model.

\begin{equation}
\label{eq:hamiltonian-tfi}
H = -h \sum_i^n \sigma_i^{x} - J\sum_{i}^n\sum_{j\in neigh(i)} \sigma_i^{z} \sigma_j^{z}.
\end{equation}

We are interested in the possible magnetic phases of the system. In the paramagnetic phase, the magnetic field $h$ dominates over the interaction $J$. The ground state is oriented in the $x$-direction and the magnetisation in the $z$-direction (and all corresponding correlations) is zero. All configurations are equally probable in this state. In the ferromagnetic phase, where $J>0$ and dominates $h$, the particles interact to align parallel to each other. The configurations where spins are parallel to each other (e.g. all spin-ups and all spin-downs) are the most probable ones.  In the antiferromagnetic phase, where $J<0$, neighbouring particles interact to align anti-parallel to each other. 
Due to the symmetry of the Ising model, the antiferromagnetic phase is equivalent to the ferromagnetic one, up to a redefinition of the directions of the spins. In particular, the transitions from paramagnetic to ferromagnetic and antiferromagnetic phases will happen at the same absolute value of $J/|h|$. In the following we consider only positive values of $h$.

We study four order parameters that are commonly used in the literature~\cite{privman1990finite,sandvik1997finite} to find the critical points. Firstly, the squared magnetisation, denoted by $M^2_F$ and shown in Equation~\ref{eq:mzf}, which shows the presence of ferromagnetism, while it is zero in the paramagnetic and antiferromagnetic phases. $M^2_F$ becoming non zero marks the transition point between the paramagnetic and ferromagnetic phase. We refer to this order parameter as the \emph{ferromagnetic magnetisation} $M^2_F$. 
\begin{equation}
\label{eq:mzf}
\begin{aligned}
M^2_F &= \frac{1}{N^2} \sum_{x \in \bm{x}}  \frac{|\Psi(x)|^{2}}{Z_{\Psi}} \left( \sum_{i=1}^{N} x_{i} \right)^2. \\
\end{aligned}
\end{equation}

Secondly, the squared magnetisation, denoted by $M^2_A$ and shown in Equation~\ref{eq:mza}, which shows the presence of antiferromagnetism, while it is zero in the paramagnetic and ferromagnetic phases. $M^2_A$ becoming non zero marks the transition point between the paramagnetic and antiferromagnetic phases. We refer to this order parameter as the \emph{antiferromagnetic magnetisation} $M^2_A$. 

\begin{equation}
\label{eq:mza}
\begin{aligned}
M^2_A &= \frac{1}{N^2} \sum_{x \in  \bm{x}} \frac{|\Psi(x)|^{2}}{Z_{\Psi}}\left( \sum_{i=1}^{N}  (-1)^{i} x_{i} \right)^2. \\
\end{aligned}
\end{equation}

Thirdly, the average ferromagnetic correlation between the particle at a given position $x_i$ to any particle $x_j$, denoted by $C_{F,i,d}$ and shown in Equation~\ref{eq:czf} where $d$ is a range of distances between two particles that we consider. This order parameter shows the inclination of the particles to be aligned with each other.  We refer to this order parameter as the \emph{ferromagnetic correlation}  $C_{F,i,d}$. 

\begin{equation}
\label{eq:czf}
\begin{aligned}
C_{F,i,d} &= \frac{1}{d-1} \sum_{x \in \bm{x}} \frac{|\Psi(x)|^{2}}{Z_{\Psi}} \left( \sum_{j=i+1}^{d+i} x_{i} x_{j} \right).  \\
\end{aligned}
\end{equation}

Finally, the average antiferromagnetic correlation between the particle at a given position $x_i$ to any particle $x_j$, denoted by $C_{A,i,d}$ and shown in Equation~\ref{eq:cza} where $d$ is a range of distances between two particles that we consider. We refer to this order parameter as the \emph{antiferromagnetic correlation} $C_{A,i,d}$.

\begin{equation}
\label{eq:cza}
\begin{aligned}
C_{A,i,d} &= \frac{1}{d-1} \sum_{x \in \bm{x}} \frac{|\Psi(x)|^{2}}{Z_{\Psi}} \left( \sum_{j=i+1}^{d+i} (-1)^{j-i} x_{i} x_{j} \right).\\
\end{aligned}
\end{equation}

The terms $(-1)^{i}$ and  $(-1)^{j-i}$ in Equations~\ref{eq:mza} and ~\ref{eq:cza}, respectively, are inserted in order to add up the magnetisation of spins that are exactly antiferromagnetically correlated. 

For one-dimensional systems, the correlation order parameters are computed from the particle at the first position ($i = 1$) to every other particle in the system ($d = N-1$). For two-dimensional systems, the correlation order parameters are computed from the first particle in the centre row (coordinate of $i = [\lfloor \sqrt{N}/2 \rfloor, 1]$) to the neighbouring particles in the same row ($ d = \sqrt{N} - 1$). For three-dimensional systems, the correlation order parameters are computed from the particle in the centre (coordinate of $i = [\lfloor \sqrt[3]{N}/2 \rfloor,1, \lfloor \sqrt[3]{N}/2 \rfloor]$) to the neighbouring particles in the same row ($d = \sqrt[3]{N} - 1$).

For the one-dimensional Ising model, in the limit of infinite size, it is exactly known that critical points are located at $J/|h| = \pm 1$~\cite{suzuki2012quantum}. The system is antiferromagnetic when $J /|h| < -1$, paramagnetic when $ -1<J / |h| < 1$ and ferromagnetic when $J / |h| >1$, which  results from the symmetry between ferromagnetic and antiferromagnetic phases. For the two-dimensional model, quantum Monte Carlo simulations~\cite{blote2002cluster} showed that the three same phases are observed, in the same order, but with critical points located at $J/|h| = \pm 0.32847$.  For the three-dimensional model, quantum Monte Carlo simulations~\cite{braiorrorrs2016phase} showed that the critical points are located at $J/|h| = \pm  0.1887$.  

\subsection{Restricted Boltzmann machine}
\label{subsec:rbm}

\begin{figure}[tb]
\centerline{\includegraphics[width=0.45\textwidth]{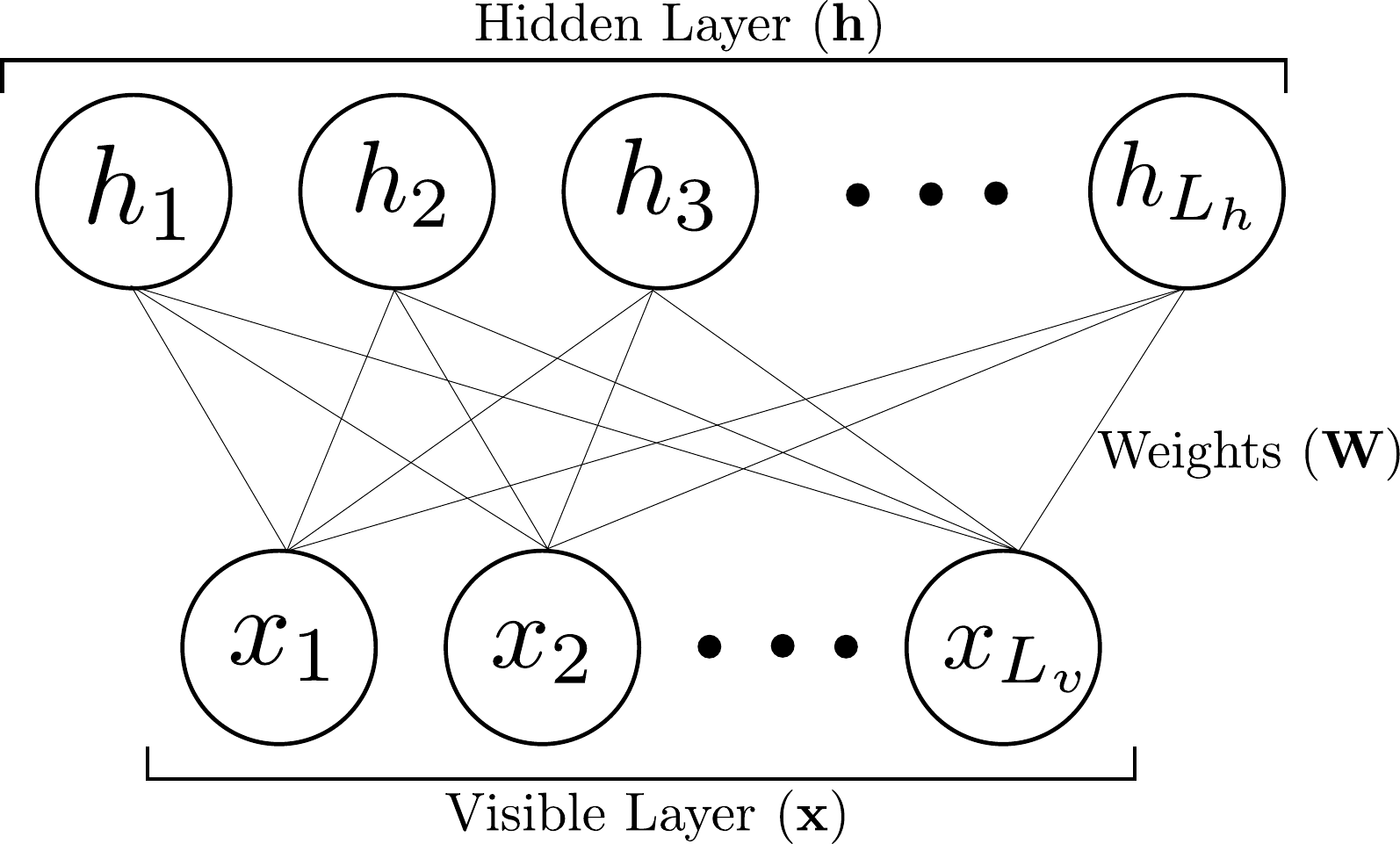}}
\caption{The structure of a restricted Boltzmann machine with $L_v$ visible nodes and $L_h$ hidden nodes. The visible layer consists of visible nodes $x_1, \dots, x_{L_v}$. The hidden layer consists of hidden nodes $h_1, \dots, h_{L_h}$. The connections between the visible and hidden layer are given by the weight matrix $\bm{W}$. \label{fig:rbm}} 
\end{figure}

A restricted Boltzmann machine is an energy-based generative model~\cite{hinton2006reducing,lecun2006tutorial}.  As shown in Fig.\ref{fig:rbm}, it consists of a visible layer $\bm{x}$ and a hidden layer $\bm{h}$. Each one of the  $N$ visible nodes $\{x_1, x_2, \cdots, x_i, \cdots, x_{N}\}$ represents the value of an input. The only design choice is the choice of the number of latent variables. It is usual to consider a multiple, $\alpha$, of the number of visible nodes. Therefore, the hidden layer consists of $\alpha \times N$ hidden nodes $\{h_1, h_2, \cdots, h_j, \cdots, h_{\alpha \times N}\}$.  The visible node $x_i$ and the hidden nodes $h_j$ are connected by the weight $W_{i,j}$. A restricted Boltzmann machine is fully described by the $N \times (\alpha \times N)$ matrix of weights. 

A restricted Boltzmann machine represents the distribution $p$ of configurations of its input layer and hidden layer as a function of its weights as given in Equation~\ref{eq:pxh}, where $Z_W$ is the normalisation constant. To get the distribution of its input layer, we marginalise $h$ from Equation~\ref{eq:pxh} to get Equation~\ref{eq:px}.  A gradient descent updating the weights can train a restricted Boltzmann machine to learn the probability distribution of a set of examples that minimises the log-likelihood whether it is supervised or unsupervised. The restricted Boltzmann machine is able to sample a configuration from this multinomial distribution. When trained with a set of example configurations, the restricted Boltzmann machine learns their distribution by minimising an energy function, which is the negative log-likelihood of the distribution. This is done by Gibbs sampling with stochastic gradient descent or contrastive divergence~\cite{hinton2012practical}.

\begin{equation}
\label{eq:pxh}
p(\mathbf{x},\mathbf{h}) = \frac{1}{Z_W} \exp\left[-\sum_{i,j}  x_i W_{i,j} h_j\right]
\end{equation}

\begin{equation}
\label{eq:px}
p(\mathbf{x}) = \frac{1}{Z_W}\prod_{j}2\cosh{\left(\sum_{i} x_i W_{ij} \right)}
\end{equation}
 
The Gibbs sampling process is as follows. From a given initial visible configuration $x$, for each hidden node $h_j$, a value is generated by sampling from the conditional probability $p(h_j \mid x)$ given in Equation~\ref{eq:hgivenv}. From this hidden configuration, for each visible node $x_i$, a value is generated by sampling from the $p(x_i \mid h)$ given in Equation~\ref{eq:vgivenh}.     

 \begin{equation}
\label{eq:hgivenv}
p(h_j = 1|x) = \mathrm{sigmoid}\left[2\left(\sum_{i}x_i W_{ij}\right)\right]
\end{equation}

 \begin{equation}
\label{eq:vgivenh}
p(x_i = 1|h) = \mathrm{sigmoid}\left[2\left(\sum_{j}W_{ij}h_j\right)\right]
\end{equation}

\subsection{Restricted Boltzmann machine neural-network quantum states}
\label{subsec:rbmnqs}

A restricted Boltzmann machine neural-network quantum state is exactly a restricted Boltzmann machine where the visible node represents one of the $N$ particles of the quantum many-body system and its value represents the value of the spin of that particle. Each node of the Restricted Boltzmann machine neural-network quantum states is a Bernoulli random variable with possible outcomes representing the values of a spin with two values, namely $-1$ or $+1$.
 
Instead of minimising the log likelihood of the distribution of training data, as it is generally the case for unsupervised energy-based machine learning models, restricted Boltzmann machine neural-network quantum states minimise the expected value of the local energy given in Equation~\ref{eq:elocrbm1}.

 \begin{align}
\label{eq:elocrbm1}
    \begin{split}
    E_{loc}(x) &= \sum_{x'} H_{x,x'}  \sqrt{\frac{p(x')} {p(x)} }\\
    &= \sum_{x'} H_{x,x'}  \sqrt{ \prod_{j}  \frac{\cosh{\left(\sum_{k} x_k^{'} W_{kj} \right)}} {\cosh{\left(\sum_{l} x_l W_{lj} \right)}}}
    \end{split}
\end{align}
       
In restricted Boltzmann machine neural-network quantum states, in order to minimise the energy of the system, leveraging the variational principle and the zero variance property, the expected value of the local energy of the configurations is minimised.  This makes the connection between the restricted Boltzmann machine neural-network quantum states and the Hamiltonian of the system it is trying to simulate. Indeed, Equation~\ref{eq:elocrbm1} is similar to Equation~\ref{eq:eloc} where the ratio of wave functions is assumed to be the same as the square root of the ratio of their norm. Here we recall that $|\Psi(x)|^2 / Z_{\Psi}$ is the probability of a configuration, and we stress here that the ground state of the Ising model can be chosen as a real and positive function, which allows us to write $\Psi(x)=\sqrt{p(x) / Z_{\Psi}}$.    
 
The unsupervised training process does not need any example. It can rely on random configurations that it generates. The iterative minimisation process alternates the Gibbs sampling of configurations, the calculation of the expected value of their local energy and stochastic gradient descent until a predefined stopping criterion is met. 
 

\section{FINDING THE QUANTUM CRITICAL POINTS}
\label{sec:transferlearning}

\subsection{Overview of the approach}
\label{subsec:algoqcp}

The approach that we consider for finding the critical points is as follows. We simulate an initial system at a selected initial parameter $J / |h|$, find its ground state and calculate the order parameter corresponding to the critical point that we are looking for. We repeat the operation increasing and decreasing the parameter with an initial step size. We are looking for an inflection point in the function of the parameter of the system that gives the value of the order parameter.  We recursively reduce the step size until we identify the inflection point. This first algorithm finds the inflection point of a system of a given size.  

The algorithm, therefore, receives the following input: the description of the system (its dimension and its size), the initial parameter $J/|h|$ of the system, the initial step size, the order parameter and the desired precision. The algorithm additionally stores the upper bound of the parameter of the system to look for the inflection point to make sure that the algorithm terminates if it does not find any inflection point. The algorithm terminates when the desired precision is reached or no inflection point is found.

We then repeat, as long as our computing resources reasonably allow, this algorithm for increasing sizes of the system. This is done to find the value of the critical point at the limit of infinite size of the system. 

We use restricted Boltzmann machine neural-network quantum states to simulate the system and calculate the order parameters. However, the repeated training of restricted Boltzmann machine neural-network quantum states for systems under different parameters and of increasing sizes is expensive. We devise three optimisations. The first, presented in Subsection~\ref{subsec:construction}, is the analytical construction of the innate restricted Boltzmann machine neural-network quantum states for a parameter deeply in the quantum phases to avoid being accidentally trapped in a local minimum. The second, presented in Subsection~\ref{subsec:transferprotocol},  is the use of transfer learning across parameters to avoid successive cold starts. The third, presented in Subsection~\ref{subsec:transfersizes}, is the use of transfer learning to larger sizes again to avoid successive cold starts. 

\subsection{Construction of innate restricted Boltzmann machine neural-network quantum states}
\label{subsec:construction}

From physical understanding, we can infer the form of the probability  distribution $| \Psi |^2$ of the configurations of a system if sufficiently deep in each phase, and construct an innate restricted Boltzmann machine neural-network quantum state that reproduces qualitatively the features of this distribution.

Several works have analytically or algorithmically constructed Boltzmann machine neural-network quantum state, e.g. \cite{carleo2018constructing}, for effective representations of quantum many-body systems. Here we use a standard restricted Boltzmann machine topology of the network, and instead we analytically evaluate its weights. 

If $J/|h|=0$, there are no interactions between spins, the system is in a deep paramagnetic phase and all the configurations are equiprobable. Putting all the weights to zero gives such distribution but forbids optimisation as all gradients are identical.  Therefore, we sample the weights from a normal distribution with zero mean and a small standard deviation.  This construction resembles the common initialisation method of the weights of a restricted Boltzmann machine~\cite{hinton2012practical}. 

If $J /|h| \to +\infty$, the interactions between particles are dominant and the system is in a deep ferromagnetic phase. The configurations where all spins are up or all spins are down are the most probable. We then construct the weights of the restricted Boltzmann machine neural-network quantum states to ensure that the probability is maximal for these two  configurations. This is achieved by setting all of the weights of each visible node to a particular hidden node to be the same and zero for the other hidden nodes. Once again, instead of using zero weights, we sample small values of the weights from a normal distribution. A similar procedure can be used for the antiferromagnetic phase when $J/|h|\to -\infty$ by setting all of the weights of each visible node to a particular hidden node to be the same but with different sign instead.

As mentioned earlier, in order to avoid being accidentally caught in a local minimum during the initial training of the first restricted Boltzmann machine neural-network quantum states  for an arbitrary initial parameter, we choose the initial parameter to be deeply in one of the phases and construct an innate restricted Boltzmann machine neural-network quantum state. We refer to this construction as RBM-NQS-I. Additionally, we refer to the restricted Boltzmann machine neural-network quantum states starting from a cold start as RBM-NQS-CS.

\subsection{Transfer learning protocol among parameters}
\label{subsec:transferprotocol}

Physically, it is expected that the wave function of systems under different but nearby values of their parameters are neighbours in the Hilbert space, although this may not be true if they are separated by a phase transition. Therefore, we expect the restricted Boltzmann machine neural-network quantum states to be similar for two systems for sufficiently nearby values of the parameters. 

Following the terminology in~\cite{yosinski2014transferable}, the base network is a trained or innate restricted Boltzmann machine neural-network quantum state for a value of the parameter of the system. The target network is a restricted Boltzmann machine neural-network quantum state for a different value of parameter with the same number of visible and hidden nodes. We can thus directly transfer the weights from the base network to the target network.

After transferring the weights, we trained the target network until it converges to a new ground state. We expect that fewer iterations are needed  for the target network to converge than it would take for a cold start initialised with a set of random weights. 

We apply this parameter transfer protocol to define an algorithm to look for the inflection point of a system of a given size. We first construct an innate restricted Boltzmann machine neural-network quantum state using RBM-NQS-I. We then calculate the order parameter value at the ground state and we iterate with this transfer learning protocol with adaptive step sizes until we locate the inflection point. We refer to this algorithm as RBM-NQS-IT.

\subsection{Transfer learning protocol to larger sizes}\label{subsec:transfersizes}

Physically, it is also expected that there is a relationship between the wave function of systems with the same parameter value but of different sizes as if they were the same system at different length scales~\cite{wilson1979problems}.  We have explored such physics-inspired transfer learning protocols in~\cite{zen2019transfer} and demonstrated their superiority over a cold start from both the effectiveness and efficiency points of view.

We want to find the critical points in the limit of infinite size. We expect the value of the parameter corresponding to the inflection points of a system of increasing finite sizes to converge asymptotically to this limit.

In our problem, this means that we need to transfer a restricted Boltzmann machine neural-network quantum state that has been optimised for a system with a certain size to another restricted Boltzmann machine neural-network quantum states with larger size and identical parameters. 

To differentiate between the two transfer learning protocols, the transfer learning protocol among phases is transferring a point within the same Hilbert space while transfer learning protocol among sizes is transferring a point across Hilbert spaces.

The base network is a restricted Boltzmann machine neural-network quantum state for a given value of the parameter of the system. The target network is a restricted Boltzmann machine neural-network quantum state for the same value but for a system of larger size. The protocol needs to leverage insights in the physics of the quantum many-body system and model. The details of the protocol are given in~\cite{zen2019transfer}.

We use this transfer learning protocol to a system of larger sizes to find the inflection point  for a series of systems of increasing sizes. Instead of starting from the same initial parameter at each size of the system, we instead start from the parameter at the inflection point  of the system of smaller size by using transfer learning protocol to larger sizes. We then find the inflection point  at the larger size. Finally, we extrapolate the value of the critical point in the limit of infinite size. We refer to this algorithm as RBM-NQS-ITT. 

We note that our method could fail because we are implementing the transfer learning for the ``hardest'' location of the parameters space, which is at the inflection point of an order parameter. Several improvements to this strategy, left for future works, could be proposed. For instance, while traversing the parameters space, we could combine the transfer learning protocol to larger sizes with the transfer learning protocol among parameters. 


\section{PERFORMANCE EVALUATION}  \label{sec:performanceeval}

The performance evaluation is threefold. We evaluate the performance of the RBM-NQS-I construction, RBM-NQS-IT for finding the inflection point for a system of a given size and RBM-NQS-ITT for finding the critical points at the limit of infinite size in Subsection~\ref{subsec:construction_eval},~\ref{subsec:size_eval} and~\ref{subsec:limit_eval}, respectively. We evaluate the effectiveness, which is the accuracy of the inflection point or the critical point, and the efficiency, which is the processing time. All of the evaluations are done for systems with open boundary conditions. 

The training of the restricted Boltzmann machine neural-network quantum states is done in an iterative manner. In each iteration, we take 10,000 samples to evaluate the local energy and its gradients. At the last iteration, we use these samples to calculate the order parameters. We update the weights using a stochastic gradient descent algorithm with RMSProp optimiser~\cite{hinton2012neural} where the initial learning rate is set to 0.001. Based on our empirical experiments, we set $\alpha = 2$ considering the efficiency and effectiveness trade-off. For RBM-NQS-CS, a random weight is sampled from a normal distribution with 0.0 mean and 0.01 standard deviation following the practical guide in~\cite{hinton2012practical}. For RBM-NQS-I, a random weight is sampled from a normal distribution with either 0.0 or 1.0 mean and 0.01 standard deviation as required by the construction. Note that the value of 1.0 whose chosen as it results in better performance after testing a range of values between 0.1 and 1.5.  

The training stops after it reaches the dynamic stopping criterion used in~\cite{zen2019transfer}, i.e. when the ratio between the standard deviation and the average of the local energy is less than 0.005 or after 30,000 iterations. Since there is randomisation involved in the training, the value reported in the paper is an average of 20 realisations of the same calculation. 

We compare this approach with the traditional methods of exact diagonalization and tensor networks. For the exact diagonalization, we use the implicitly restarted Arnoldi method to find the eigenvalues and eigenvectors~\cite{lehoucq1998arpack}. Our computational resources only allow us to compute exact diagonalization up to 20 particles. For the tensor network method, we use the matrix product states algorithm~\cite{schollwock2011} with a bond dimension up to 1000. Both of the methods run only once since there is no randomisation involved.

The existing code of RBM-NQS is implemented in C++ with support for Message Passing Interface under a library named NetKet~\cite{netket:2019}. We ported the code into TensorFlow library~\cite{abadi2016tensorflow} for a significant speedup with the graphics processing units.  All of the experiments run on an NVIDIA DGX-1 server equipped with NVIDIA Tesla V100 graphics processing units with 640 tensor cores, 5120 CUDA cores, and 16GB memory.  

For the algorithm to find the inflection point, we choose the initial step size as $1.0$ and we divide the step size by $10$ after one iteration. The algorithm stops when the precision is $10^{-3}$. To calculate the gradient, we use second-order accurate central differences. 

\subsection{Evaluation of innate restricted Boltzmann machine neural-network quantum states}
\label{subsec:construction_eval}

The performance evaluation of RBM-NQS-I deeply in each phase is twofold. First, we construct RBM-NQS-I without training and evaluate them. Second, we fine-tune the RBM-NQS-I until it reaches the stopping criterion and evaluate them. We evaluate the effectiveness and efficiency by comparing the value of the energy and the order parameters and by comparing the iterations needed for the training until it reached the stopping criterion with RBM-NQS-CS, respectively.

We choose $J / |h| = 0.0$, $J / |h| = 3.0$ and $J / |h| = -3.0$ for the cases of deep paramagnetic, ferromagnetic and antiferromagnetic phases, respectively. In the ferromagnetic and antiferromagnetic case, the weights are sampled from a normal distribution with either 0.0 or 1.0 mean and 0.01 standard deviation as prescribed in Subsection~\ref{subsec:construction}.     
Table~\ref{tab:cons_128_0.0}, Table~\ref{tab:cons_128_3.0} and Table~\ref{tab:cons_128_-3.0}  show the evaluation of the RBM-NQS-CS and RBM-NQS-I where the size of the system is $128$ and parameter of the system $J/|h| = 0.0$, $J/|h| = 3.0$ and  $J/|h| = -3.0$, respectively. 

In the case of a deep paramagnetic phase ($J / |h| = 0.0$ in Table~\ref{tab:cons_128_0.0}), we observe that both the energy and the order parameter for both the RBM-NQS-CS and RBM-NQS-I without training are very close to the result of the tensor network method. When we train the RBM-NQS-I, it stops directly because it already reaches the stopping criterion. The value of the energy and order parameter are not exactly the same as the tensor network value due to the noise introduced in the weights and from the sampling process.

In the case of a deep ferromagnetic phase ($J / |h| = 3.0$ in Table~\ref{tab:cons_128_3.0}) and antiferromagnetic phase ($J / |h| = -3.0$ in Table~\ref{tab:cons_128_-3.0}), we observe that the results of the RBM-NQS-I are closer to the result of the tensor network method and need less iterations to converge to the stopping criterion than RBM-NQS-CS. The energy and the order parameters of the RBM-NQS-I without training is quite far from the result of the tensor network method. We hypothesise that this is because $J/|h| = 3$ or $J/|h| = -3$ is not deep enough in the ferromagnetic phase. To evaluate the hypothesis, we comparatively evaluate RBM-NQS-I on system with 16 particles for $J/|h| = 3$, $J/|h| = 5$ and $J/|h| = 10$. We observe that the relative error of the energy with the exact diagonalisation of RBM-NQS-I for $J/|h| = 3$, $J/|h| = 5$ and $J/|h| = 10$ is $0.216$, $0.101$ and $0.047$, respectively. This means that RBM-NQS-I is better when it is deeper in the corresponding phase.

For RBM-NQS-CS, we observe that the value of the order parameters are very far even though the energy is closer to the result of the tensor network method. This means that the training of RBM-NQS-CS remains in a local minimum and the restricted Boltzmann machine does not converge to the ground state. 

Table~\ref{tab:cons_4_0.0}, Table~\ref{tab:cons_4_3.0} and Table~\ref{tab:cons_4_-3.0} in Appendix~\ref{sec:construction_eval_2d_app}  show the evaluation of the RBM-NQS-CS and RBM-NQS-I for two-dimensional system where the size of the system is $4 \times 4$ and parameter of the system $J/|h| = 0.0$, $J/|h| = 3.0$ and  $J/|h| = -3.0$, respectively.     Table~\ref{tab:cons_222_0.0}, Table~\ref{tab:cons_222_3.0} and Table~\ref{tab:cons_222_-3.0} in Appendix~\ref{sec:construction_eval_3d_app}  show the evaluation of the RBM-NQS-CS and RBM-NQS-I for three-dimensional system where the size of the system is $2 \times 2 \times 2$ and parameter of the system $J/|h| = 0.0$, $J/|h| = 3.0$ and  $J/|h| = -3.0$, respectively. We have chosen such system sizes so as to be able to compare the neural network quantum state results to exact diagonalization calculations.

We see similar trends for both the result of the two-dimensional and three-dimensional systems  with those of the one-dimensional system.

In the case of a deep paramagnetic phase ($J / |h| = 0.0$ in Table~\ref{tab:cons_4_0.0}  in Appendix~\ref{sec:construction_eval_2d_app} and Table~\ref{tab:cons_222_0.0}  in Appendix~\ref{sec:construction_eval_3d_app} ), the results of the RBM-NQS-CS and RBM-NQS-I are very close to the result of the exact diagonalization method. Therefore, no further training is needed.  

In the case of a deep ferromagnetic phase ($J / |h| = 3.0$ in Table~\ref{tab:cons_4_3.0}  in Appendix~\ref{sec:construction_eval_2d_app}  and Table~\ref{tab:cons_222_3.0}  in Appendix~\ref{sec:construction_eval_3d_app}) and antiferromagnetic phase ($J / |h| = -3.0$ in Table~\ref{tab:cons_4_-3.0}  in Appendix~\ref{sec:construction_eval_2d_app} and Table~\ref{tab:cons_222_-3.0}  in Appendix~\ref{sec:construction_eval_3d_app}), we see that the results of RBM-NQS-I are closer to those of the exact diagonalization calculations than RBM-NQS-CS before training. We also observe that the constructed RBM-NQS-I needs less iterations to converge to the stopping criterion than RBM-NQS-CS. Therefore, in two-dimensional and three-dimensional systems, we conclude that RBM-NQS-I is more effective than RBM-NQS-CS before training. However, they are equally effective after training but RBM-NQS-I is more efficient than RBM-NQS-CS.

To conclude, we showed that RBM-NQS-I on one-dimensional, two-dimensional and three-dimensional systems deep in each phase is more effective and efficient than RBM-NQS-CS. Furthermore, further training is not needed in the case of a deep paramagnetic phase (i.e. $J / |h| = 0$). Therefore, from this point forward, we choose $J/|h| = 0$ as our initial parameter in our algorithm for finding the inflection point.     

\begin{table*}
\begin{center}
 {\caption{The performance evaluation of the RBM-NQS-CS and RBM-NQS-I for one-dimensional system in Ising model where the system size is $128$ and parameter of the system $J/|h| = 0.0$. The reported value is average value over 20 realisations. The value inside the parentheses is the standard deviation.  }\label{tab:cons_128_0.0}}
\begin{tabular}{@{}cccccc@{}}
\toprule
\multicolumn{1}{l}{\multirow{2}{*}{}} & \multicolumn{2}{c}{\textbf{Without training}} & \multicolumn{2}{c}{\textbf{With training}}  & \multirow{2}{*}{\textbf{Tensor network}} \\ \cmidrule(lr){2-5}
\multicolumn{1}{l}{}                  & \textbf{RBM-NQS-CS}  & \textbf{RBM-NQS-I}  & \textbf{RBM-NQS-CS} & \textbf{RBM-NQS-I} &                                          \\ \midrule
\textbf{Energy}                       & -127.9799 (0.0029)   & -127.9799 (0.0029)     & -127.9799 (0.0029)  & -127.9799 (0.0029)    & -128.00000                               \\ \midrule
\textbf{$\bm{M_F^2}$}                    & 0.0079 (0.0001)      & 0.0079 (0.0001)        & 0.0079 (0.0001)     & 0.0079 (0.0001)       & 0.00781                                  \\ \midrule
\textbf{$\bm{M_A^2}$}                    & 0.0078 (0.0001)      & 0.0078 (0.0001)        & 0.0078 (0.0001)     & 0.0078 (0.0001)       & 0.00781                                  \\ \midrule
\textbf{$\bm{C_{F,i,d}}$}                    & 0.0003 (0.0007)      & 0.0003 (0.0007)        & 0.0003 (0.0007)     & 0.0003 (0.0007)       & 0.00000                                  \\ \midrule
\textbf{$\bm{C_{A,i,d}}$}                    & -0.0003 (0.0007)     & -0.0003 (0.0007)       & -0.0003 (0.0007)    & -0.0003 (0.0007)      & 0.00000                                  \\ \midrule
\textbf{Iterations}                    & -                    & -                      & 0                   & 0                     & -                                        \\ \bottomrule
\end{tabular}
\end{center}
\end{table*}

\begin{table*}
\begin{center}
 {\caption{The performance evaluation of RBM-NQS-CS and RBM-NQS-I for one-dimensional system in Ising model where the system size is $128$ and parameter of the system $J/|h| = 3.0$. The reported value is average value over 20 realisations. The value inside the parentheses is the standard deviation.  }\label{tab:cons_128_3.0}}
\begin{tabular}{@{}cccccc@{}}
\toprule
\multicolumn{1}{l}{\multirow{2}{*}{}} & \multicolumn{2}{c}{\textbf{Without training}} & \multicolumn{2}{c}{\textbf{With training}}    & \multirow{2}{*}{\textbf{Tensor network}} \\ \cmidrule(lr){2-5}
\multicolumn{1}{l}{}                  & \textbf{RBM-NQS-CS}  & \textbf{RBM-NQS-I}  & \textbf{RBM-NQS-CS} & \textbf{RBM-NQS-I} &                                          \\ \midrule
\textbf{Energy}                       & -127.9061 (0.2577)   & -217.5726 (0.3152)     & -372.2911 (4.7748)    & -391.7046 (0.0182)    & -391.91198                               \\ \midrule
\textbf{$\bm{M_F^2}$}         & 0.0078 (0.0001)      & 0.2934 (0.0009)        & 0.0981 (0.1041)       & 0.9658 (0.0005)       & 0.96980                                  \\ \midrule
\textbf{$\bm{M_A^2}$}         & 0.0078 (0.0001)      & 0.0056 (0.0001)        & 0.0006 (0.0001)       & 0.0003 (0.0000)       & 0.00023                                  \\ \midrule
\textbf{$\bm{C_{F,i,d}}$}         & -0.0002 (0.0011)     & 0.2897 (0.0059)        & 0.0360 (0.2928)       & 0.9362 (0.0051)       & 0.92871                                  \\ \midrule
\textbf{$\bm{C_{A,i,d}}$}         & 0.0001 (0.0010)      & -0.0020 (0.0009)       & -0.0025 (0.0032)      & -0.0072 (0.0002)      & -0.00704                                 \\ \midrule
\textbf{Iterations}                    & -                    & -                      & 1621.7250 (1271.7007) & 20.8500 (0.4770)      &  -                                     \\ \bottomrule
\end{tabular}
\end{center}
\end{table*}

\begin{table*}
\begin{center}
 {\caption{The performance evaluation of the RBM-NQS-CS and RBM-NQS-I for one-dimensional system in Ising model where the system size is $128$ and parameter of the system $J/|h| = -3.0$. The reported value is average value over 20 realisations. The value inside the parentheses is the standard deviation.  }\label{tab:cons_128_-3.0}}
\begin{tabular}{@{}cccccc@{}}
\toprule
\multicolumn{1}{l}{\multirow{2}{*}{}} & \multicolumn{2}{c}{\textbf{Without training}} & \multicolumn{2}{c}{\textbf{With training}}   & \multirow{2}{*}{\textbf{Tensor network}} \\ \cmidrule(lr){2-5}
\multicolumn{1}{l}{}                  & \textbf{RBM-NQS-CS}  & \textbf{RBM-NQS-I}  & \textbf{RBM-NQS-CS} & \textbf{RBM-NQS-I} &                                          \\ \midrule
\textbf{Energy}                       & -128.0531 (0.2487)   & -217.8645 (0.4201)     & -371.8311 (5.1623)   & -391.7077 (0.0165)    & -391.91198                               \\ \midrule
\textbf{$\bm{M_F^2}$}                    & 0.0078 (0.0001)      & 0.0055 (0.0001)        & 0.0006 (0.0001)      & 0.0003 (0.0000)       &  0.00023                               \\ \midrule
\textbf{$\bm{M_A^2}$}                    & 0.0078 (0.0001)      & 0.2939 (0.0010)        & 0.1276 (0.1287)      & 0.9658 (0.0004)       &  0.96980                                   \\ \midrule
\textbf{$\bm{C_{F,i,d}}$}                    & 0.0004 (0.0009)      & -0.0018 (0.0008)       & -0.0022 (0.0031)     & -0.0072 (0.0002)      & -0.00704                                \\ \midrule
\textbf{$\bm{C_{A,i,d}}$}                    & 0.0003 (0.0008)      & 0.2874 (0.0050)        & 0.0176 (0.3385)      & 0.9349 (0.0058)       &   0.92871                                 \\ \midrule
\textbf{Iterations}                    & -                    & -                      & 1331.6000 (720.1564) & 20.8500 (0.4770)      & -                                        \\ \bottomrule
\end{tabular}
\end{center}
\end{table*}

\subsection{Finding inflection point for a system of a given size}
\label{subsec:size_eval}
We evaluate the performance of the algorithm for finding the inflection point for a system of a given size with RBM-NQS-IT.

The performance evaluation is twofold. We first provide an analysis by plotting the values of the order parameter as a function of the parameter $J/|h|$.  We then evaluate the inflection point for each system's size and compare the value to other traditional methods to compare its effectiveness.

First, we plot the value of the order parameters as a function of $J/|h|$. We use RBM-NQS-CS and RBM-NQS-IT to compute the order parameter at the ground state of each point in the space of the parameter of the system. For efficiency, we compare the time needed for all of the computation. 

For one-dimensional systems, we calculate the order parameters for $J/|h|$ within the range $[-2,2]$ with $0.1$ intervals and for systems with size $N=\{8,16,32,64,128\}$. For two-dimensional systems, we calculate the order parameters for $J/|h|$ within the range $[-1,1]$ with $0.1$ intervals and for systems with sizes $N=\{2\times 2,4 \times 4,8\times 8,16\times 16\}$. For three-dimensional systems, we calculate the order parameters for $J/|h|$ within the range $[-1,1]$ with $0.1$ intervals and for systems with sizes $N=\{2\times 2 \times 2,3 \times 3 \times 3,4 \times 4 \times 4,5\times 5\times 5,6\times 6 \times 6\}$.

\begin{figure*}[!htb]
\centerline{\includegraphics[width=\textwidth]{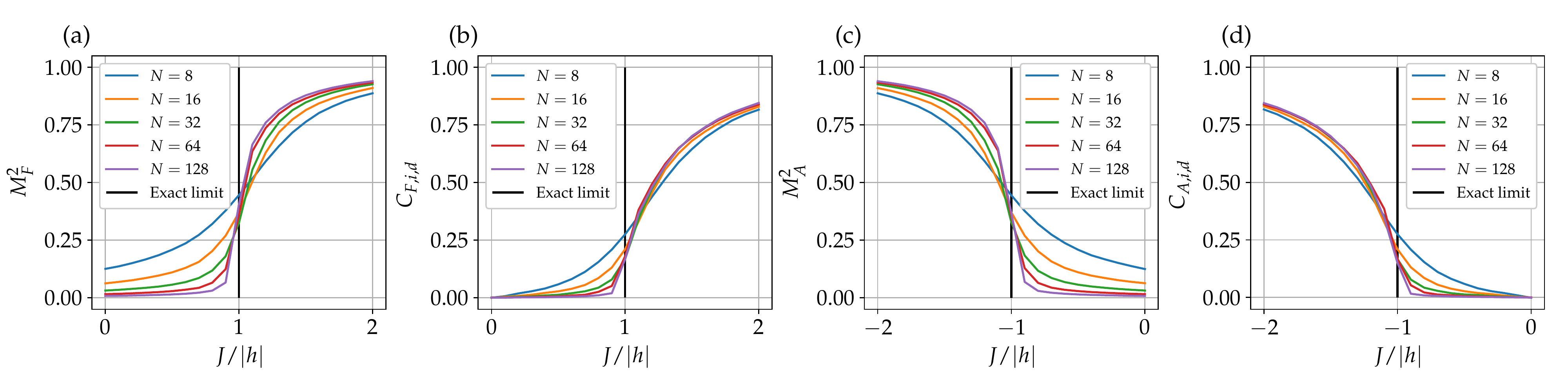}}
\caption{The value of order parameters with RBM-NQS-IT for one-dimensional systems for $J/|h|$ within the range $[-2,0]$ and $[0,2]$, for antiferromagnetic and ferromagnetic order parameters respectively, with $0.1$ intervals and for system with size $N=\{8,16,32,64,128\}$. (a), (b), (c) and (d) shows the ferromagnetic magnetisation $M_F^2$, ferromagnetic correlation $C_{F,i,d}$, antiferromagnetic magnetisation $M_A^2$ and antiferromagnetic correlation $C_{A,i,d}$, respectively. The exact critical point at the limit of infinite size is at $J/|h|= \pm 1$~\cite{suzuki2012quantum}. \label{fig:all-mf-1d} }
\end{figure*}

Figure~\ref{fig:all-mf-1d} shows the value of the order parameters for one-dimensional systems with RBM-NQS-IT. In the limit of infinite size, there should be an abrupt change of the derivative at the critical point and the value of the order parameter should change from 0 to an increasing function. We observe that the change in the derivative of the order parameter is more abrupt as we increase the size of the system. Similarly, we also observe that the value of the order parameter are closer to zero in one phase and closer to a function of the distance from the critical point in the other phase as we increase the size of the system. Figure~\ref{fig:all-mf-1d-dmrg} in Appendix~\ref{sec:size_eval_app_dmrg}  shows the result for the tensor network method and it shows a similar trend as observed in Figure~\ref{fig:all-mf-1d}. 

We observe that the weights of RBM-NQS-IT do not change so drastically throughout the space of the parameters of the model. We expect that this is because the order parameters that we study behave smoothly close to the transition even for the large system sizes we consider. Video animations of the weights of RBM-NQS-IT for one realisation of a system of size $32$ with the ferromagnetic $M^2_F$ and the antiferromagnetic $M^2_A$ magnetisation order parameters and for $J/|h|$ from $-3$ to $3$ with $0.1$ intervals, are available online\footnote{Video for ferromagnetic magnetisation: \url{https://youtu.be/OSKBC8Fm2r4}, video for antiferromagnetic  magnetisation: \url{https://youtu.be/kTEzdVfVNMA}.}.   

\begin{figure*}[!htb]
\centerline{\includegraphics[width=\textwidth]{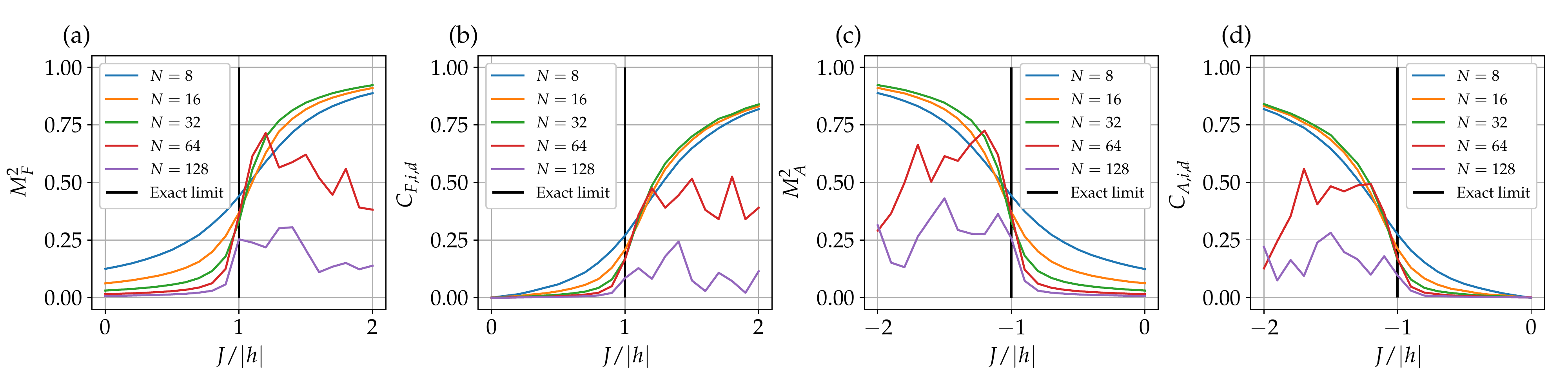}}
\caption{The value of order parameters with RBM-NQS-CS for one-dimensional systems for $J/|h|$ within the range $[-2,0]$ and $[0,2]$, for antiferromagnetic and ferromagnetic order parameters respectively, with $0.1$ intervals and for system with size $N=\{8,16,32,64,128\}$. (a), (b), (c) and (d) shows the ferromagnetic magnetisation $M_F^2$, ferromagnetic correlation $C_{F,i,d}$, antiferromagnetic magnetisation $M_A^2$ and antiferromagnetic correlation $C_{A,i,d}$, respectively. The exact critical point at the limit of infinite size is at $J/|h|=\pm 1$~\cite{suzuki2012quantum}. \label{fig:all-mf-1d-cs}}
\end{figure*}

Figure~\ref{fig:all-mf-1d-cs} shows the value of the order parameters for one-dimensional systems with RBM-NQS-CS. We observe that every order parameters fails to get to the correct value before or after the inflection point for systems with size $64$ and $128$. This is possibly due to the network being trapped in a local minimum and more parameter tuning is needed. Therefore, RBM-NQS-IT is more effective than RBM-NQS-CS.

Figure~\ref{fig:all-mf-2d} and Figure~\ref{fig:all-mf-2d-cs} in Appendix~\ref{sec:size_eval_app_2d} show the value of the order parameters for two-dimensional systems with RBM-NQS-IT and RBM-NQS-CS, respectively. Figure~\ref{fig:all-mf-3d} and Figure~\ref{fig:all-mf-3d-cs} in Appendix~\ref{sec:size_eval_app_3d} show the value of the order parameters for three-dimensional systems with RBM-NQS-IT and RBM-NQS-CS, respectively. We see a similar trend as the result of the one-dimensional model.  In two-dimensional systems, RBM-NQS-CS remains in a local minimum for a size $16\times 16$. We note that, in three dimensions, RBM-NQS-CS performs well even for a system of size $6\times 6 \times 6$. This may be due to the fact that correlations are not as strong in a system with larger connectivity, i.e. each site is coupled to more sites.         

It takes approximately 10 minutes to compute one realisation of a system with the size of 128 particles with RBM-NQS-IT, where by a realisation we mean the computation for $J/|h|$ within the range $[-2,2]$ with values spaced by intervals of $0.1$.  Meanwhile, RBM-NQS-CS takes approximately 5 hours and the tensor network method that we have implemented takes approximately 60 hours. While this is not a fair comparison, we show here that the restricted Boltzmann machine neural-network quantum states leveraging graphics processing units give very good computing times. Furthermore, given the reduced number of iterations required, RBM-NQS-IT boosts the speed even further. 

Next, we evaluate the inflection point for each system's size. We evaluate the performance on one-dimensional systems from $N = 8$ and doubling each time until $N = 128$.  For two-dimensional systems, we start from $N = 2 \times 2$ and doubling each time until $N = 16 \times 16$. For three-dimensional systems, we instead start from $N = 2 \times 2 \times 2$ and increment the size of the system by one until $N = 6 \times 6 \times 6$. For three-dimensional systems, we do not use the transfer learning protocol across sizes since we do not double the size of the system. 

We comparatively evaluate the effectiveness and efficiency of RBM-NQS-CS and RBM-NQS-IT. To evaluate the effectiveness, we compare the value of the inflection point at each size of the system with the tensor network method~\cite{schollwock2011} and exact diagonalization for one-dimensional systems. For two-dimensional and three-dimensional systems, we only compare with exact diagonalization. 

Table~\ref{tab:qcp_1d_mf} and Table~\ref{tab:qcp_1d_ma} show the value of the inflection point  for different sizes of the system of one-dimensional, two-dimensional and three-dimensional systems with RBM-NQS-CS, RBM-NQS-IT and tensor network method with ferromagnetic magnetisation $M_F^2$ and  antiferromagnetic magnetisation $M_A^2$ order parameter, respectively. 

In Table~\ref{tab:qcp_1d_mf} and Table~\ref{tab:qcp_1d_ma}, we observe that RBM-NQS-CS performs the worst overall since the value of the inflection point  is far from both the tensor network and exact diagonalization methods, especially in system of large size. It is particularly unstable in a one-dimensional system with 64 and 128 particles, as shown by a very large standard deviation. 

We observe that the tensor network method is closer to the exact diagonalization method for systems of small size than RBM-NQS-IT. We see that both the inflection point for RBM-NQS-IT and tensor network converge towards $J/|h| = \pm 1$, the exact critical point at the infinite size limit~\cite{suzuki2012quantum}. 

In two-dimensional and three-dimensional system, both the results of RBM-NQS-CS and RBM-NQS-IT are close to the exact diagonalization method. However, RBM-NQS-IT is closer to the exact diagonalization method result than RBM-NQS-CS by a small margin. We believe that the performance of RBM-NQS-CS and RBM-NQS-IT is similar because of the small sizes considered, which were chosen so as to be able to compare to exact diagonalization results.

Table~\ref{tab:qcp_1d_cf} and Table~\ref{tab:qcp_1d_ca} in Appendix~\ref{sec:size_eval_app_quant}  show the value of the inflection point  for different sizes of the system of one-dimensional, two-dimensional and three-dimensional systems with RBM-NQS-CS, RBM-NQS-IT and tensor network method with ferromagnetic correlation $C_{F,i,d}$ and antiferromagnetic correlation $C_{A,i,d}$ order parameter, respectively. It shows similar trends as those observed for the magnetisation order parameters.

To evaluate the efficiency, we compare the time needed to detect the inflection point. It takes approximately 5 hours for one realisation to find the inflection point  for innate RBM-NQS-IT for a system with a size of 128 particles. However, the absolute variance of the inflection point  is relatively small, around 0.001. Therefore, in practice, one run suffices.
Even though the RBM-NQS-CS takes approximately less than 1 hour, it is unstable and gives a wrong value  for the inflection point. The tensor network method takes approximately 20 hours to find the inflection point.

\begin{table}
\begin{center}
 {\caption{The value of the inflection point  for one-, two- and three-dimensional systems of given sizes with RBM-NQS-CS, RBM-NQS-IT, tensor network and exact diagonalization method with ferromagnetic magnetisation $M_F^2$ order parameter. The value inside the parentheses is the standard deviation.}\label{tab:qcp_1d_mf}\label{tab:qcp_2d_mf}\label{tab:qcp_3d_mf}}

\begin{tabular}{ccccc}
\toprule
\textbf{\begin{tabular}[c]{@{}c@{}}System \\ size\end{tabular}} & \textbf{\begin{tabular}[c]{@{}c@{}}RBM-NQS-CS\end{tabular}} & \textbf{\begin{tabular}[c]{@{}c@{}}RBM-NQS-IT\end{tabular}} & \textbf{\begin{tabular}[c]{@{}c@{}}Tensor\\ network\end{tabular}} & \textbf{\begin{tabular}[c]{@{}c@{}}Exact\\ diag.\end{tabular}} \\ \midrule
$8$                                                               & 1.114 (0.009)                                                          & 1.105 (0.006)                                                                 & 1.11                                                              & 1.109                                                                    \\ \midrule
$16$                                                             & 1.007 (0.008)                                                          & 1.040 (0.005)                                                                 & 1.08                                                              & 1.090                                                                    \\ \midrule
$32$                                                              & 1.011 (0.009)                                                          & 1.013 (0.001)                                                                 & 1.05                                                              & -                                                             \\ \midrule
$64$                                                              & 1.004 (0.009)                                                          & 1 (0.001)                                                                     & 1.02                                                              & -                                                            \\ \midrule
$128$                                                             & 0.646 (0.38)                                                           & 1 (0.001)                                                                     & 1.01                                                              & -                                                            \\ \midrule \midrule
$2\times2$                                                    & 0.662 (0.04)                                                           & 0.673 (0.05)                                                                 & -  & 0.69                                                                     \\ \midrule
$4\times4$                                                    &  0.5 (0.0)                                                              & 0.501 (0.003)                                                               & -    & 0.51     
                                                           \\ \midrule\midrule
$2\times2\times2$                                                    &  0.505 (0.012)                                                           & 0.502 (0.004)                                                                 & -  & 0.527                                                                     
\\ \bottomrule
\end{tabular}         
\end{center}

\end{table}

\begin{table}
\begin{center}
 {\caption{The value of the inflection point  for one-, two- and three-dimensional systems of given sizes with RBM-NQS-CS, RBM-NQS-IT, tensor network and exact diagonalization method with antiferromagnetic magnetisation $M_A^2$ order parameter. The value inside the parentheses is the standard deviation.}\label{tab:qcp_1d_ma}\label{tab:qcp_2d_ma}\label{tab:qcp_3d_ma}}

\begin{tabular}{ccccc}
\toprule
\textbf{\begin{tabular}[c]{@{}c@{}}System \\ size\end{tabular}} & \textbf{\begin{tabular}[c]{@{}c@{}}RBM-NQS-CS\end{tabular}} & \textbf{\begin{tabular}[c]{@{}c@{}}RBM-NQS-IT\end{tabular}} & \textbf{\begin{tabular}[c]{@{}c@{}}Tensor\\ network\end{tabular}} & \textbf{\begin{tabular}[c]{@{}c@{}}Exact\\ diag.\end{tabular}} \\ \midrule
$8$                                                             & -1.072 (0.05)       & -1.10 (0.005)       & -1.12                                                             & -1.109                                                         \\ \midrule
$16$                                                            & -1.012 (0.01)       & -1.035 (0.006)      & -1.08                                                             & -1.090                                                         \\ \midrule
$32$                                                            & -1.011 (0.01)       & -1.010 (0.004)      & -1.05                                                             & -                                                              \\ \midrule
$64$                                                            & -1.108 (0.36)       & -1.004 (0.003)      & -1.02                                                             & -                                                              \\ \midrule
128                                                             & -0.912 (0.21)       & -1.002 (0.002)      & -1.01                                                             & -                                                              \\ \midrule\midrule
$2 \times 2$                                                    & -0.624 (0.08)       & -0.656 (0.05)       & -                                                                 & -0.69                                                          \\ \midrule
$4 \times 4$                                                    & -0.5 (0.0)          & -0.5 (0.0)          & -                                                                 & -0.51                                                          \\ \midrule\midrule
$2 \times 2 \times 2$                                           & -0.502 (0.002)      & -0.502 (0.001)      & -                                                                 & -0.527                                            
                                                                 
\\ \bottomrule
\end{tabular}         
\end{center}

\end{table}

\subsection{Finding quantum critical points at the limit of infinite size}
\label{subsec:limit_eval}

We evaluate the effectiveness of RBM-NQS-ITT for finding the critical points at the limit of infinite size. We use the $(L,2)-$tiling protocol defined in~\cite{zen2019transfer} for the transfer learning protocol to larger sizes by transferring the parameters at the inflection point of a smaller size system to a larger one. 

The performance evaluation is twofold. We first provide an analysis by plotting the values of the order parameter as a function of the parameter $J/|h|$, which has been done in Subsection~\ref{subsec:size_eval}. We then provide an evaluation by fitting the value of the inflection point  at each size of the system to show towards which value it converges in the infinite-size limit.    

We observe in Figure~\ref{fig:all-mf-1d} that with RBM-NQS-IT, for all order parameters, the  inflection point converges toward $\pm 1.0$, which is the exact critical point at the limit of infinite size~\cite{suzuki2012quantum}, as we increase the size of the system.

For two-dimensional and three-dimensional systems, we observe similar trends as those observed in one-dimensional systems. For two-dimensional systems, we observe in Figure~\ref{fig:all-mf-2d} and~\ref{fig:all-mf-2d-cs} in Appendix~\ref{sec:construction_eval_2d_app} that the inflection point for RBM-NQS-IT and RBM-NQS-CS, respectively, is converging toward the value of the critical point at the limit of infinite size $\pm  0.32847$ obtained with a quantum Monte Carlo method~\cite{blote2002cluster}. Similarly to one-dimensional systems, even though RBM-NQS-CS may remain trapped in a local minimum close to $\pm  0.32847$ for a system with size $16 \times 16$, the inflection point is still close to  $\pm  0.32847$. For three-dimensional systems, we observe in Figure~\ref{fig:all-mf-3d} and~\ref{fig:all-mf-3d-cs} in Appendix~\ref{sec:construction_eval_3d_app} that the inflection point for RBM-NQS-IT and RBM-NQS-CS, respectively, is converging toward the value of the critical point at the limit of infinite size $\pm  0.1887$ obtained with a quantum Monte Carlo method~\cite{braiorrorrs2016phase}. 

We evaluate the value of the critical point at the limit of infinite size by extrapolating a series of inflection points at increasing system sizes as a function of the size of the system. We fit a function of the form $f(N) = a + b\;N^{c}$ with non-linear least squares, where $a$, $b$ and $c$ are the function parameters. The constraint of the parameter is $b > 0$ and $c < 0$ for ferromagnetic order parameters and $b < 0$ and $c < 0$ for antiferromagnetic order parameters. The value of $a$ approximates the value of the critical point at the limit of infinite size. We exclude the RBM-NQS-CS from this evaluation since we have shown in the previous sections that RBM-NQS-IT effectiveness is better.

\begin{figure}[tb]
\centerline{\includegraphics[width=0.45\textwidth]{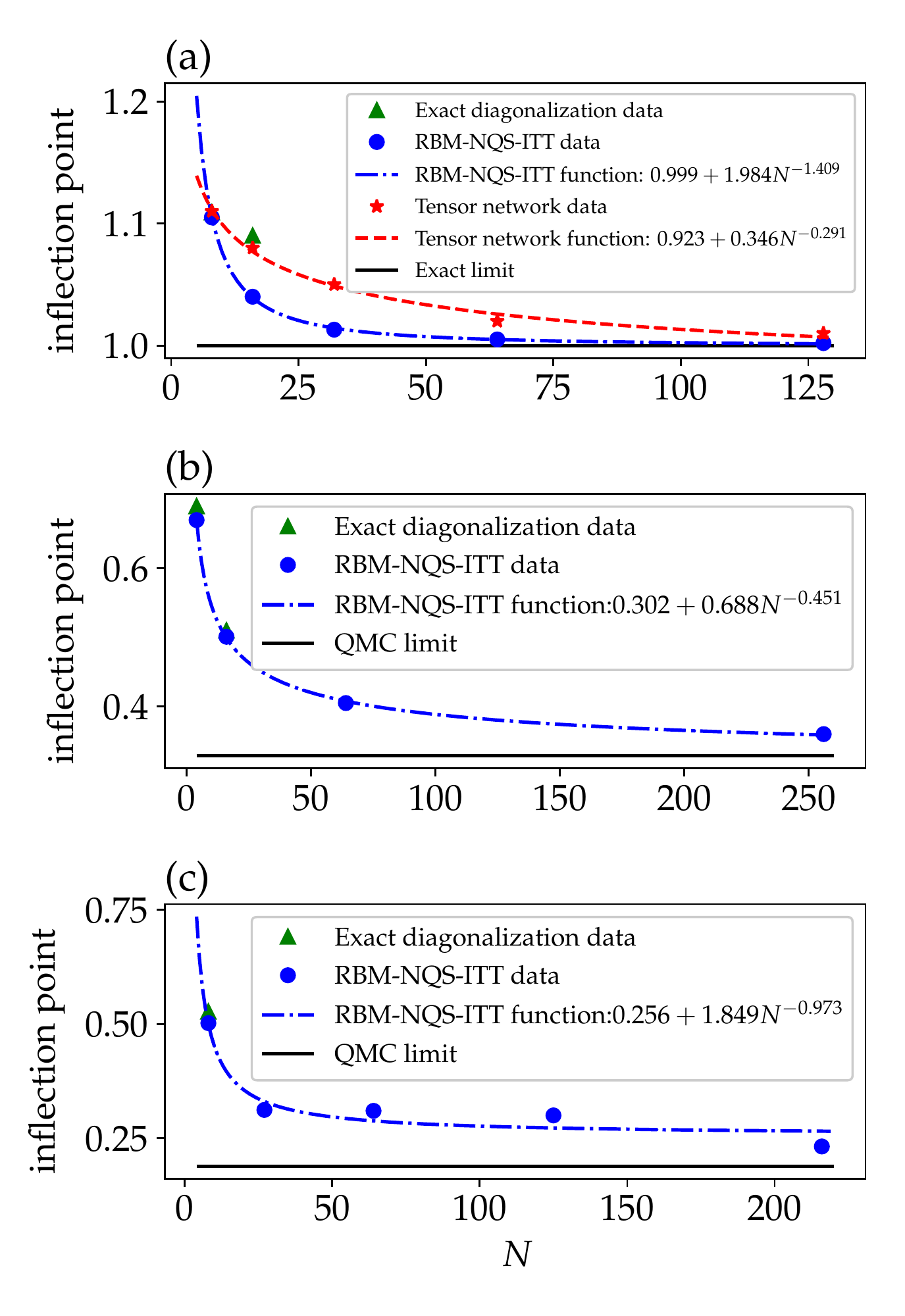}}
\caption{The evaluation of the critical point at the limit of infinite size by fitting the inflection points  as a function of the size of the system in one- (a), two- (b), three- (c) dimensional models.  We use the ferromagnetic magnetisation $M_F^2$ to find the critical point. The critical point at the limit of infinite size is at $J/|h|=1$~\cite{suzuki2012quantum} in one-dimensional system, $J/|h|=0.32847$~\cite{blote2002cluster} in two-dimensional system and $J/|h|=0.1887$~\cite{braiorrorrs2016phase} in three-dimensional system.  \label{fig:limit-1d-mzf}\label{fig:limit-2d-mzf}} 
\end{figure}

Figure~\ref{fig:limit-1d-mzf} (a) shows the evaluation of the critical point at the limit of infinite size by fitting the inflection points  as a function of the size of the system in the one-dimensional model with ferromagnetic magnetisation $M^2_F$ order parameter. We compare the result of RBM-NQS-ITT with the tensor network method. The value of $a$ is $0.999$ and $0.923$ for RBM-NQS-ITT and the tensor network method, respectively.

Figure~\ref{fig:limit-1d-mzf} (b) shows the same evaluation for systems in two dimensions. The value of $a$ on RBM-NQS-ITT is $0.302$, which is close to the value $0.32847$ based on quantum Monte Carlo method~\cite{blote2002cluster}. Figure~\ref{fig:limit-1d-mzf} (c) shows the same evaluation for systems in three-dimensions. The value of $a$ on RBM-NQS-ITT is $0.256$, which is sizeably different to the value $0.1887$ based on quantum Monte Carlo method~\cite{blote2002cluster}. We expect that we need systems of larger sizes for three-dimensional system to better estimate the critical point.

\begin{figure}[tb]
\centerline{\includegraphics[width=0.45\textwidth]{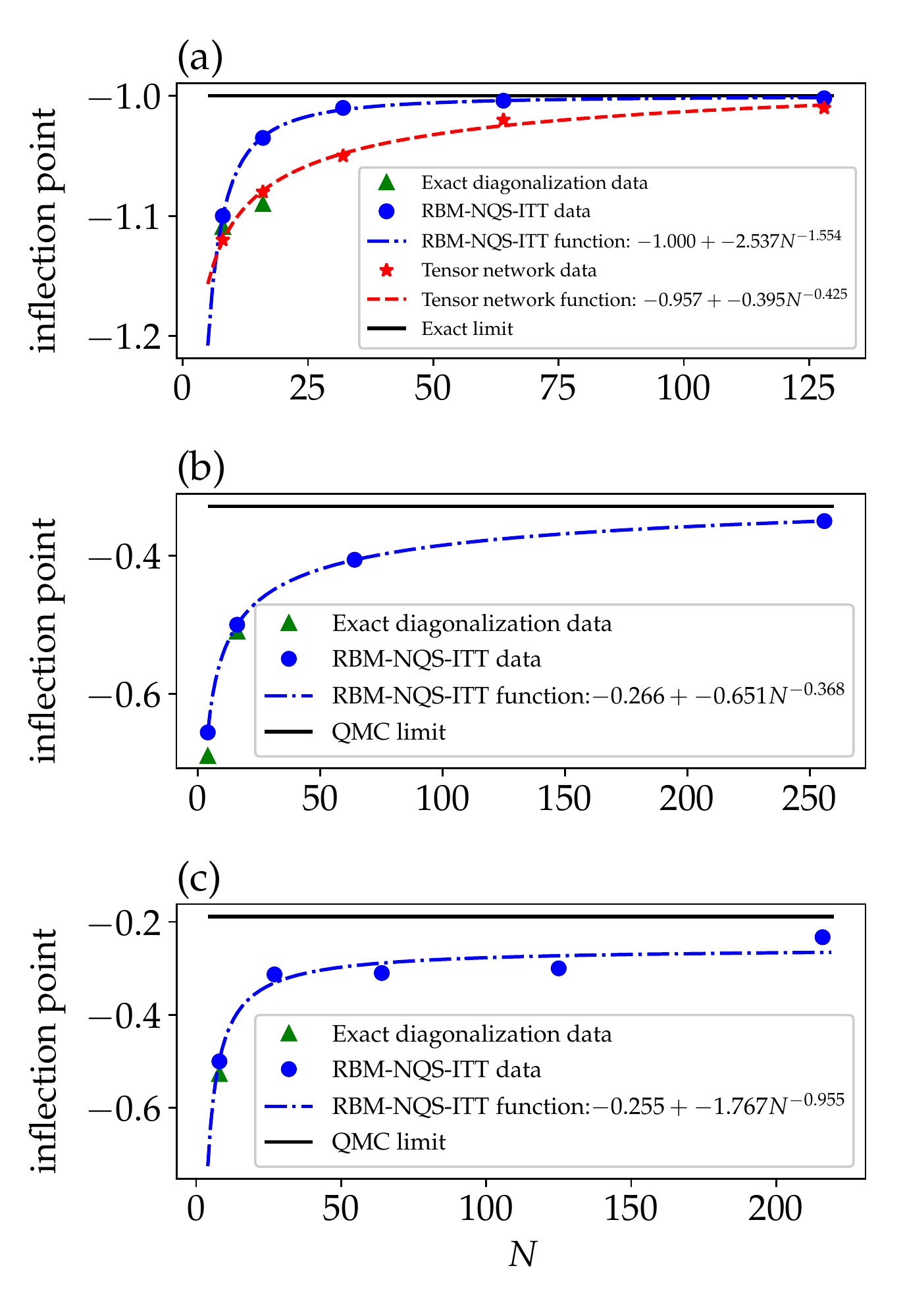}}
\caption{The evaluation of the critical point at the limit of infinite size by fitting the inflection points  as a function of the size of the system in one- (a), two- (b), three- (c) dimensional models.  We use the antiferromagnetic magnetisation $M_A^2$ to find the critical point. The critical point at the limit of infinite size is at $J/|h|=-1$~\cite{suzuki2012quantum} in one-dimensional system, $J/|h|=-0.32847$~\cite{blote2002cluster} in two-dimensional system and $J/|h|=-0.1887$~\cite{braiorrorrs2016phase} in three-dimensional system.  \label{fig:limit-1d-mza}\label{fig:limit-2d0mza}} 
\end{figure}

Figure~\ref{fig:limit-1d-mza} shows the same evaluation using the antiferromagnetic magnetisation $M^2_A$ order parameter. We observe similar trends as those observed in the ferromagnetic one except that the value at the limit for two-dimensional system is further than the ferromagnetic one to the quantum Monte Carlo limit. The value of $a$ is $-0.266$ for the antiferromagnetic phase where the critical point at the limit of infinite size is at $-0.32847$.  

Figure~\ref{fig:limit-1d-czf} and~\ref{fig:limit-1d-cza} in Appendix~\ref{sec:limit_eval_app} show the same evaluation using the correlation order parameters. We observe similar trends as those observed for the magnetisation order parameters. However, using the ferromagnetic correlation $C_{F,i,d}$ order parameter on one-dimensional systems, we see that the tensor network method gives better correlations, as the extrapolated critical point is closer to the exact limit than RBM-NQS-ITT.

\section{CONCLUSION} \label{sec:conclusion}

We have proposed an approach to finding quantum critical points with innate restricted Boltzmann machine neural-network quantum states and transfer learning protocols. We applied the proposed approach to one-, two- and three-dimensional Ising models and in the limit of infinite size.

We have empirically and comparatively shown that our proposed approach is more effective and efficient than cold start approaches, which start from a network with randomly initialised parameters. It is also more efficient than traditional approaches. Furthermore, we have shown that we can estimate the value of the quantum critical point at the infinite size limit with transfer learning protocol to larger sizes as proposed in~\cite{zen2019transfer}.

A natural extension to this work is the study of the quantum critical exponents, which describe the behaviour of the order parameters close to the phase transitions. 
We also would like to further explore the opportunities to analytically and algebraically construct neural-network quantum states. Such approaches may be used to devise solutions to other problems such as characterisation of properties of different quantum many-body systems, the study of their time evolution, as well as the study of quantum few-body systems.

Mathematically, our proposed transfer learning protocols are done inside and across the Hilbert space of the wave function. Therefore, we also would like to pursue an application of the reverse where machine learning algorithms, especially energy-based generative model, are explained in terms of Hilbert space and physical systems.

\ack We acknowledge C. Guo and Supremacy Future Technologies for support on the matrix product states simulations. This work is partially funded by the National University of Singapore, the French Ministry of European and Foreign Affairs and the French Ministry of Higher Education, Research and Innovation under the Merlion programme, project ``Deep Quantum''. We acknowledge support from the Singapore Ministry of Education, Singapore Academic Research Fund Tier II (project MOE2018-T2-2-142). The experiments reported in this article are performed on the infrastructure of Singapore National Supercomputing Centre\footnote{\url{https://nscc.sg}} and are funded under project ``Computing the Deep Quantum''. 

\bibliography{ecai}

\appendix
\onecolumn

\section{Evaluation of innate restricted Boltzmann machine neural-network quantum states for two-dimensional systems}
\label{sec:construction_eval_2d_app}

Table~\ref{tab:cons_4_0.0}, Table~\ref{tab:cons_4_3.0} and Table~\ref{tab:cons_4_-3.0} shows the evaluation of the RBM-NQS-CS and RBM-NQS-I for two-dimensional system where the size of the system is $4 \times 4$ and parameter of the system $J/|h| = 0.0$, $J/|h| = 3.0$ and  $J/|h| = -3.0$, respectively. 

\begin{table*}[!htb]
\begin{center}
 {\caption{The performance evaluation of RBM-NQS-CS and RBM-NQS-I for two-dimensional system in Ising model where the system size is $4\times4$ and parameter of the system $J/|h| = 0.0$. The reported value is average value over 20 realisations. The value inside the parentheses is the standard deviation.  }\label{tab:cons_4_0.0}}
\begin{tabular}{@{}cccccc@{}}
\toprule
\multicolumn{1}{l}{\multirow{2}{*}{}} & \multicolumn{2}{c}{\textbf{Without training}} & \multicolumn{2}{c}{\textbf{With training}}  & \multirow{2}{*}{\textbf{\begin{tabular}[c]{@{}c@{}}Exact \\ Diagonalization\end{tabular}}} \\ \cmidrule(lr){2-5}
\multicolumn{1}{l}{}                  & \textbf{RBM-NQS-CS}  & \textbf{RBM-NQS-I}  & \textbf{RBM-NQS-CS} & \textbf{RBM-NQS-I} &                                          \\ \midrule
\textbf{Energy}                       & -16.0000 (0.0001)    & -16.0000 (0.0001)      & -16.0000 (0.0001)   & -16.0000 (0.0001)     & -16.0000                                                                                   \\ \midrule
\textbf{$\bm{M_F^2}$}                    & 0.0624 (0.0007)      & 0.0624 (0.0007)        & 0.0624 (0.0007)     & 0.0624 (0.0007)       & 0.0625                                                                                     \\ \midrule
\textbf{$\bm{M_A^2}$}                    & 0.0624 (0.0009)      & 0.0624 (0.0009)        & 0.0624 (0.0009)     & 0.0624 (0.0009)       & 0.0625                                                                                     \\ \midrule
\textbf{$\bm{C_{F,i,d}}$}                    & 0.2503 (0.0041)      & 0.2503 (0.0041)        & 0.2503 (0.0041)     & 0.2503 (0.0041)       & 0.2500                                                                                     \\ \midrule
\textbf{$\bm{C_{A,i,d}}$}                    & 0.2490 (0.0037)      & 0.2490 (0.0037)        & 0.2490 (0.0037)     & 0.2490 (0.0037)       & 0.2500                                                                                     \\ \midrule
\textbf{Iterations}                    & -                    & -                      & 0                   & 0                     & -                                                                                          \\ \bottomrule
\end{tabular}
\end{center}
\end{table*}

\begin{table*}[!htb]
\begin{center}
 {\caption{The performance evaluation of RBM-NQS-CS and RBM-NQS-I for two-dimensional system in Ising model where the system size is $4\times4$ and parameter of the system $J/|h| = 3.0$. The reported value is average value over 20 realisations. The value inside the parentheses is the standard deviation.  }\label{tab:cons_4_3.0}}
\begin{tabular}{@{}cccccc@{}}
\toprule
\multicolumn{1}{l}{\multirow{2}{*}{}} & \multicolumn{2}{c}{\textbf{Without training}} & \multicolumn{2}{c}{\textbf{With training}}  & \multirow{2}{*}{\textbf{\begin{tabular}[c]{@{}c@{}}Exact \\ Diagonalization\end{tabular}}} \\ \cmidrule(lr){2-5}
\multicolumn{1}{l}{}                  & \textbf{RBM-NQS-CS}  & \textbf{RBM-NQS-I}  & \textbf{RBM-NQS-CS} & \textbf{RBM-NQS-I} &                                          \\ \midrule
\textbf{Energy}                       & -15.9808 (0.1652)    & -52.1501 (0.1441)      & -72.9401 (0.0035)   & -72.9397 (0.0039)     & -72.9455                                                                                   \\ \midrule
\textbf{$\bm{M_F^2}$}                    & 0.0625 (0.0008)      & 0.6064 (0.0023)        & 0.9856 (0.0005)     & 0.9861 (0.0005)       & 0.9860                                                                                     \\ \midrule
\textbf{$\bm{M_A^2}$}                    & 0.0624 (0.0007)      & 0.0263 (0.0003)        & 0.0010 (0.0000)     & 0.0009 (0.0000)       & 0.0009                                                                                     \\ \midrule
\textbf{$\bm{C_{F,i,d}}$}                    & 0.2483 (0.0042)      & 0.6863 (0.0056)        & 0.9924 (0.0009)     & 0.9924 (0.0009)       & 0.9934                                                                                     \\ \midrule
\textbf{$\bm{C_{A,i,d}}$}                    & 0.2520 (0.0045)      & 0.1042 (0.0036)        & 0.0022 (0.0004)     & 0.0024 (0.0006)       & 0.0017                                                                                     \\ \midrule
\textbf{Iterations}                    & -                    & -                      & 180.4000 (4.8311)   & 126.0500 (1.7741)     & -                                                                                          \\ \bottomrule
\end{tabular}
\end{center}
\end{table*}

\begin{table*}[!htb]
\begin{center}
 {\caption{The performance evaluation of RBM-NQS-CS and RBM-NQS-I for two-dimensional system in Ising model where the system size is $4\times4$ and parameter of the system $J/|h| = -3.0$. The reported value is average value over 20 realisations. The value inside the parentheses is the standard deviation.  }\label{tab:cons_4_-3.0}}
\begin{tabular}{@{}cccccc@{}}
\toprule
\multicolumn{1}{l}{\multirow{2}{*}{}} & \multicolumn{2}{c}{\textbf{Without training}} & \multicolumn{2}{c}{\textbf{With training}}  & \multirow{2}{*}{\textbf{\begin{tabular}[c]{@{}c@{}}Exact \\ Diagonalization\end{tabular}}} \\ \cmidrule(lr){2-5}
\multicolumn{1}{l}{}                  & \textbf{RBM-NQS-CS}  & \textbf{RBM-NQS-I}  & \textbf{RBM-NQS-CS} & \textbf{RBM-NQS-I} &                                          \\ \midrule
\textbf{Energy}                       & -16.0528 (0.1601)    & -52.1049 (0.1821)      & -72.9396 (0.0032)   & -72.9403 (0.0043)     & -72.9455                                                                                   \\ \midrule
\textbf{$\bm{M_F^2}$}                    & 0.0625 (0.0007)      & 0.0263 (0.0004)        & 0.0010 (0.0000)     & 0.0009 (0.0000)       & 0.0009                                                                                     \\ \midrule
\textbf{$\bm{M_A^2}$}                    & 0.0627 (0.0009)      & 0.6059 (0.0026)        & 0.9857 (0.0005)    & 0.9861 (0.0006)       & 0.9860                                                                                     \\ \midrule
\textbf{$\bm{C_{F,i,d}}$}                    & 0.2510 (0.0031)      & 0.1052 (0.0026)        & 0.0022 (0.0004)     & 0.0022 (0.0004)       & 0.0017                                                                                     \\ \midrule
\textbf{$\bm{C_{A,i,d}}$}                    & 0.2492 (0.0050)      & 0.6852 (0.0031)        & 0.9924 (0.0008)     & 0.9923 (0.0009)       & 0.9934                                                                                     \\ \midrule
\textbf{Iterations}                    & -                    & -                      & 179.5500 (3.6807)   & 125.1500 (1.3143)     & -                                                                                          \\ \bottomrule
\end{tabular}
\end{center}
\end{table*}

\section{Evaluation of innate restricted Boltzmann machine neural-network quantum states for three-dimensional systems}
\label{sec:construction_eval_3d_app}
Table~\ref{tab:cons_222_0.0}, Table~\ref{tab:cons_222_3.0} and Table~\ref{tab:cons_222_-3.0}  shows the evaluation of the RBM-NQS-CS and RBM-NQS-I for three-dimensional system where the size of the system is $2 \times 2 \times 2$ and parameter of the system $J/|h| = 0.0$, $J/|h| = 3.0$ and  $J/|h| = -3.0$, respectively.

\begin{table*}[!htb]
\begin{center}
 {\caption{The performance evaluation of RBM-NQS-CS and RBM-NQS-I for three-dimensional system in Ising model where the system size is $2\times 2\times 2$ and parameter of the system $J/|h| = 0.0$. The reported value is average value over 20 realisations. The value inside the parentheses is the standard deviation.  }\label{tab:cons_222_0.0}}
\begin{tabular}{@{}cccccc@{}}
\toprule
\multicolumn{1}{l}{\multirow{2}{*}{}} & \multicolumn{2}{c}{\textbf{Without training}} & \multicolumn{2}{c}{\textbf{With training}}  & \multirow{2}{*}{\textbf{\begin{tabular}[c]{@{}c@{}}Exact \\ Diagonalization\end{tabular}}} \\ \cmidrule(lr){2-5}
\multicolumn{1}{l}{}                  & \textbf{RBM-NQS-CS}  & \textbf{RBM-NQS-I}  & \textbf{RBM-NQS-CS} & \textbf{RBM-NQS-I} &                                          \\ \midrule
\textbf{Energy}                       & -8.0000 (0.0000)     & -8.0000 (0.0000)       & -8.0000 (0.0000)    & -8.0000 (0.0000)      & -8                                                                                         \\ \midrule
\textbf{$\bm{M_F^2}$}                    & 0.1244 (0.0022)      & 0.1244 (0.0022)        & 0.1244 (0.0022)     & 0.1244 (0.0022)       & 0.125                                                                                      \\ \midrule
\textbf{$\bm{M_A^2}$}                    & 0.1250 (0.0028)      & 0.1250 (0.0028)        & 0.1250 (0.0028)     & 0.1250 (0.0028)       & 0.125                                                                                      \\ \midrule
\textbf{$\bm{C_{F,i,d}}$}                    & 0.4999 (0.0059)      & 0.4999 (0.0059)        & 0.4999 (0.0059)     & 0.4999 (0.0059)       & 0.5                                                                                        \\ \midrule
\textbf{$\bm{C_{A,i,d}}$}                    & 0.5000 (0.0059)      & 0.5000 (0.0059)        & 0.5000 (0.0059)     & 0.5000 (0.0059)       & 0.5                                                                                        \\ \midrule
\textbf{Iterations}                    & -                    & -                      & 0.0000 (0.0000)     & 0.0000 (0.0000)       & - \\ \bottomrule
\end{tabular}
\end{center}
\end{table*}

\begin{table*}[!htb]
\begin{center}
 {\caption{The performance evaluation of RBM-NQS-CS and RBM-NQS-I for three-dimensional system in Ising model where the system size is $2\times 2\times 2$ and parameter of the system $J/|h| = 3.0$. The reported value is average value over 20 realisations. The value inside the parentheses is the standard deviation.  }\label{tab:cons_222_3.0}}
\begin{tabular}{@{}cccccc@{}}
\toprule
\multicolumn{1}{l}{\multirow{2}{*}{}} & \multicolumn{2}{c}{\textbf{Without training}} & \multicolumn{2}{c}{\textbf{With training}}  & \multirow{2}{*}{\textbf{\begin{tabular}[c]{@{}c@{}}Exact \\ Diagonalization\end{tabular}}} \\ \cmidrule(lr){2-5}
\multicolumn{1}{l}{}                  & \textbf{RBM-NQS-CS}  & \textbf{RBM-NQS-I}  & \textbf{RBM-NQS-CS} & \textbf{RBM-NQS-I} &                                          \\ \midrule
\textbf{Energy}                       & -7.9931 (0.1337)     & -32.8919 (0.0726)      & -36.4436 (0.0022)   & -36.4445 (0.0015)     & -36.4451                                                                                   \\ \midrule
\textbf{$\bm{M_F^2}$}                    & 0.1254 (0.0022)      & 0.8416 (0.0026)        & 0.9887 (0.0007)     & 0.9884 (0.0004)       & 0.9891                                                                                     \\ \midrule
\textbf{$\bm{M_A^2}$}                    & 0.1252 (0.0018)      & 0.0225 (0.0004)        & 0.0016 (0.0001)     & 0.0017 (0.0001)       & 0.0015                                                                                     \\ \midrule
\textbf{$\bm{C_{F,i,d}}$}                    & 0.5023 (0.0056)      & 0.9094 (0.0026)        & 0.9936 (0.0014)     & 0.9938 (0.0008)       & 0.9938                                                                                     \\ \midrule
\textbf{$\bm{C_{A,i,d}}$}                    & 0.4977 (0.0056)      & 0.0906 (0.0026)        & 0.0064 (0.0014)     & 0.0062 (0.0008)       & 0.0062                                                                                     \\ \midrule
\textbf{Iterations}                    & -                    & -                      & 274.6000 (7.4726)   & 171.8000 (2.2716)     & -                                                                                              \\ \bottomrule
\end{tabular}
\end{center}
\end{table*}

\begin{table*}[!htb]
\begin{center}
 {\caption{The performance evaluation of RBM-NQS-CS and RBM-NQS-I for three-dimensional system in Ising model where the system size is $2\times 2\times 2$ and parameter of the system $J/|h| = -3.0$. The reported value is average value over 20 realisations. The value inside the parentheses is the standard deviation.  }\label{tab:cons_222_-3.0}}
\begin{tabular}{@{}cccccc@{}}
\toprule
\multicolumn{1}{l}{\multirow{2}{*}{}} & \multicolumn{2}{c}{\textbf{Without training}} & \multicolumn{2}{c}{\textbf{With training}}  & \multirow{2}{*}{\textbf{\begin{tabular}[c]{@{}c@{}}Exact \\ Diagonalization\end{tabular}}} \\ \cmidrule(lr){2-5}
\multicolumn{1}{l}{}                  & \textbf{RBM-NQS-CS}  & \textbf{RBM-NQS-I}  & \textbf{RBM-NQS-CS} & \textbf{RBM-NQS-I} &                                          \\ \midrule
\textbf{Energy}                       & -8.0406 (0.0767)     & -32.8799 (0.0948)      & -36.4422 (0.0032)   & -36.4431 (0.0019)     & -36.4451                                                                                   \\ \midrule
\textbf{$\bm{M_F^2}$}                    & 0.1243 (0.0013)      & 0.0227 (0.0005)        & 0.0016 (0.0001)     & 0.0017 (0.0001)       & 0.0015                                                                                     \\ \midrule
\textbf{$\bm{M_A^2}$}                    & 0.1250 (0.0013)      & 0.8412 (0.0028)        & 0.9887 (0.0005)     & 0.9880 (0.0010)       & 0.9891                                                                                     \\ \midrule
\textbf{$\bm{C_{F,i,d}}$}                    & 0.4996 (0.0031)      & 0.0908 (0.0040)        & 0.0061 (0.0008)     & 0.0075 (0.0007)       & 0.0062                                                                                     \\ \midrule
\textbf{$\bm{C_{A,i,d}}$}                    & 0.5004 (0.0031)      & 0.9092 (0.0040)        & 0.9939 (0.0008)     & 0.9925 (0.0007)       & 0.9938                                                                                     \\ \midrule
\textbf{Iterations}                    & -                    & -                      & 273.5000 (5.3898)   & 171.5000 (1.9621)     & -                                                                                          \\ \bottomrule
\end{tabular}
\end{center}
\end{table*}

\section{Analysis of the order parameter for a system of a given size with tensor network method}
\label{sec:size_eval_app_dmrg}
\begin{figure*}[!htb]
\centerline{\includegraphics[width=\textwidth]{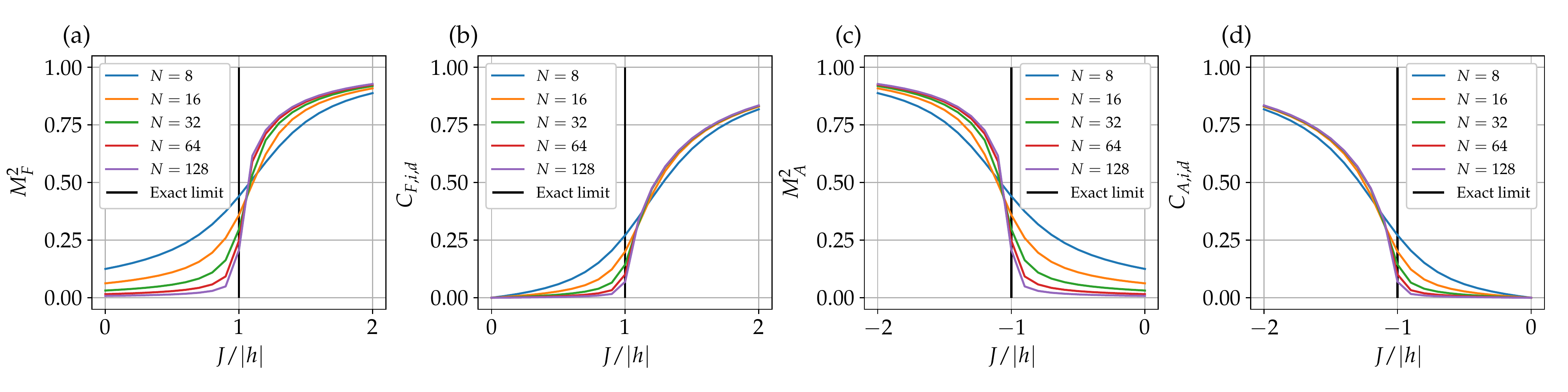}}
\caption{The value of order parameters with the tensor network method for one-dimensional systems for $J/|h|$ within the range $[-2,0]$ and $[0,2]$, for antiferromagnetic and ferromagnetic order parameters respectively,  with $0.1$ intervals and for system with size $N=\{8,16,32,64,128\}$. (a), (b), (c) and (d) shows the ferromagnetic magnetisation $M_F^2$, ferromagnetic correlation $C_{F,i,d}$, antiferromagnetic magnetisation $M_A^2$ and antiferromagnetic correlation $C_{A,i,d}$  order parameter, respectively. The exact critical point at the limit of infinite size is at $J/|h|= \pm 1$~\cite{suzuki2012quantum}. \label{fig:all-mf-1d-dmrg}}
\end{figure*}

Figure~\ref{fig:all-mf-1d-dmrg} shows the value of the order parameters for a one-dimensional system with the tensor network method. We calculate the order parameters for $J/|h|$ within the $[-2,0]$ and $[0,2]$, for antiferromagnetic and ferromagnetic order parameters respectively, with $0.1$ intervals and for systems with sizes $N=\{8,16,32,64,128\}$.

\section{Analysis of the order parameter for a system of a given size for two dimensional systems}
\label{sec:size_eval_app_2d}

\begin{figure*}[!htb]
\centerline{\includegraphics[width=\textwidth]{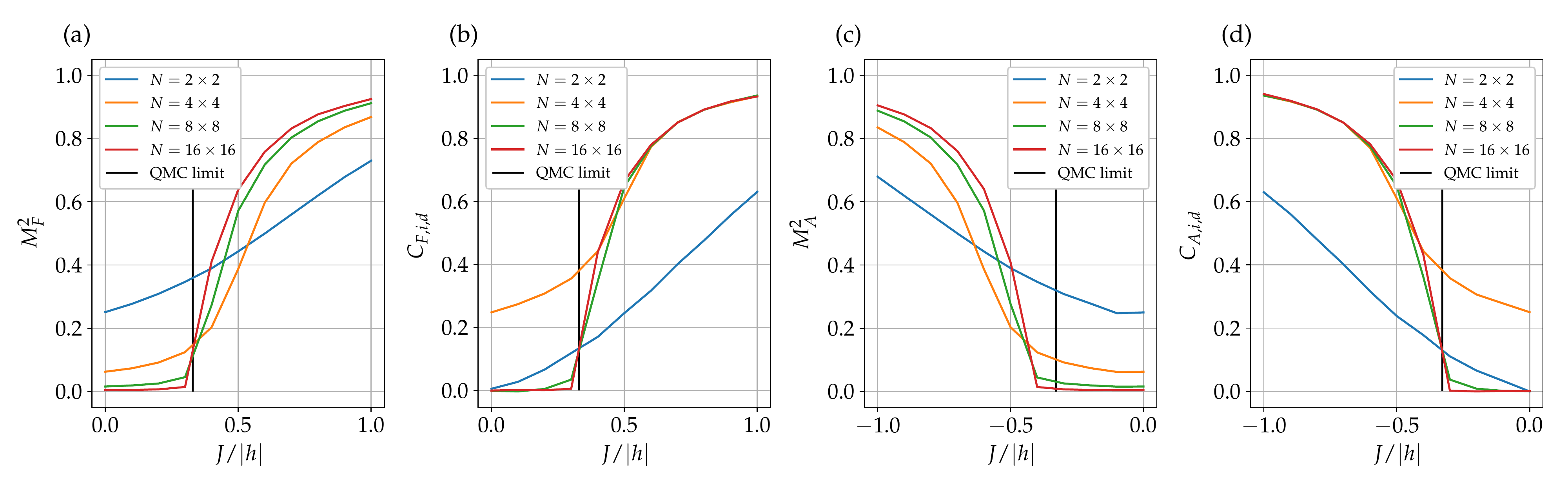}}
\caption{The value of order parameters with RBM-NQS-IT for two-dimensional systems for $J/|h|$ within the range $[-1,0]$ and $[0,1]$, for antiferromagnetic and ferromagnetic order parameters respectively, with $0.1$ intervals and for system with size $N=\{2\times2, 4\times4, 8\times8, 16\times16\}$. (a), (b), (c) and (d) shows the ferromagnetic magnetisation $M_F^2$, ferromagnetic correlation $C_{F,i,d}$, antiferromagnetic magnetisation $M_A^2$ and antiferromagnetic correlation $C_{A,i,d}$  order parameter, respectively. The critical point at the limit of infinite size is at  $\pm 0.32847$ based on quantum Monte Carlo method (QMC)~\cite{blote2002cluster}.  \label{fig:all-mf-2d}}
\end{figure*}

\begin{figure*}[!htb]
\centerline{\includegraphics[width=\textwidth]{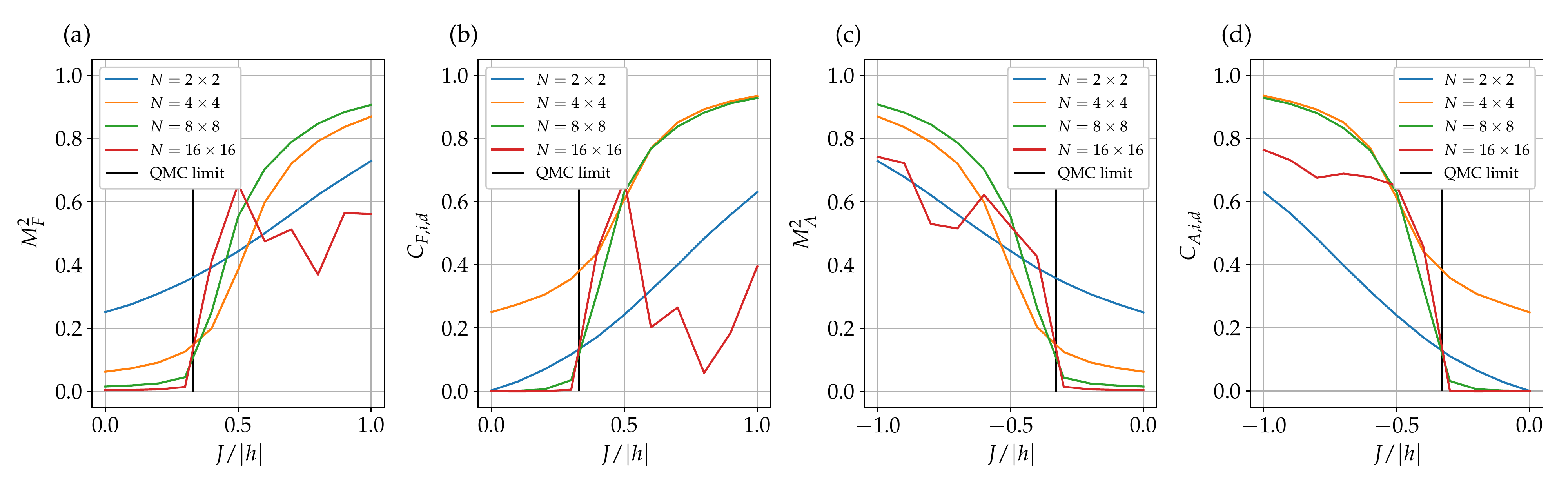}}
\caption{The value of order parameters with RBM-NQS-CS for two-dimensional systems for $J/|h|$ within the range $[-1,0]$ and $[0,1]$, for antiferromagnetic and ferromagnetic order parameters respectively, with $0.1$ intervals and for system with size $N=\{2\times2, 4\times4, 8\times8, 16\times16\}$. (a), (b), (c) and (d) shows the ferromagnetic magnetisation $M_F^2$, ferromagnetic correlation $C_{F,i,d}$, antiferromagnetic magnetisation $M_A^2$ and antiferromagnetic correlation $C_{A,i,d}$ order parameter, respectively. The critical point at the limit of infinite size is at  $\pm 0.32847$ based on quantum Monte Carlo method (QMC)~\cite{blote2002cluster}.  \label{fig:all-mf-2d-cs}}
\end{figure*}

Figure~\ref{fig:all-mf-2d} and Figure~\ref{fig:all-mf-2d-cs}  shows the value of the order parameters for a two-dimensional system with RBM-NQS-IT and RBM-NQS-CS, respectively. We calculate the order parameters for $J/|h|$ within the range $[-1,0]$ and $[0,1]$, for antiferromagnetic and ferromagnetic order parameters respectively,  with $0.1$ intervals and for systems with sizes $N=\{2\times 2,4 \times 4,8\times 8,16\times 16\}$.

\section{Analysis of the order parameter for a system of a given size for three dimensional systems}
\label{sec:size_eval_app_3d}

\begin{figure*}[!htb]
\centerline{\includegraphics[width=\textwidth]{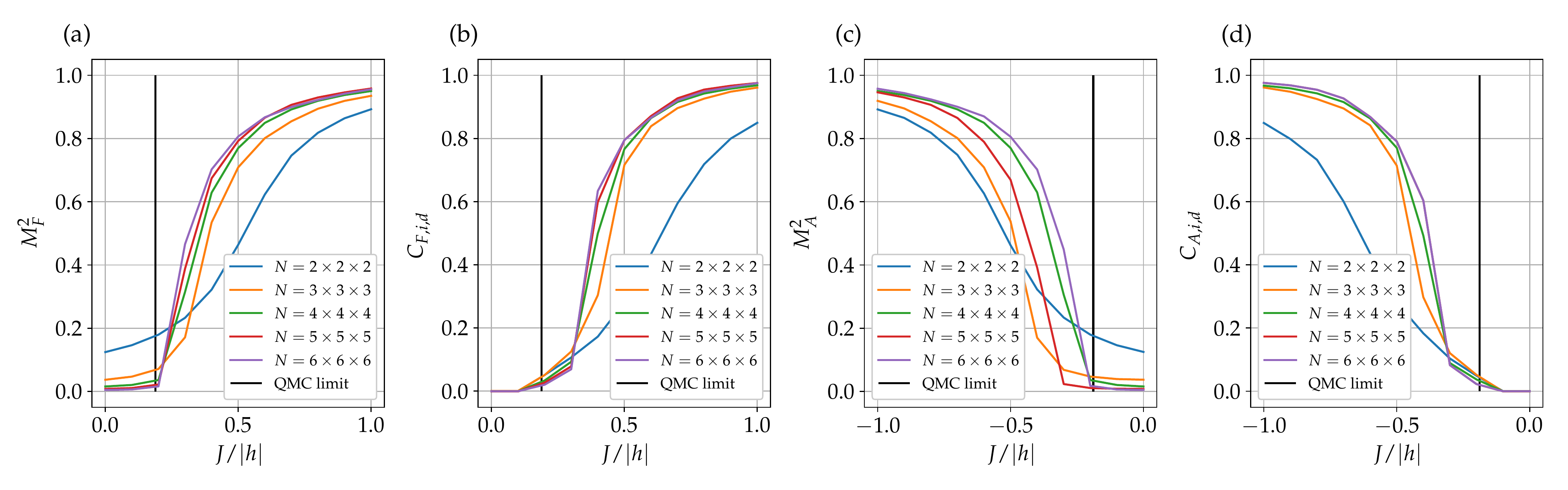}}
\caption{The value of order parameters with RBM-NQS-IT for three-dimensional systems for $J/|h|$ within the range $[-1,0]$ and $[0,1]$, for antiferromagnetic and ferromagnetic order parameters respectively, with $0.1$ intervals and for system with size $N=\{2\times2\times2, 3\times3\times3, 4\times4\times4, 5\times5\times5, 6\times6\times6 \}$. (a), (b), (c) and (d) shows the ferromagnetic magnetisation $M_F^2$, ferromagnetic correlation $C_{F,i,d}$, antiferromagnetic magnetisation $M_A^2$ and antiferromagnetic correlation $C_{A,i,d}$ order parameter, respectively. The critical point at the limit of infinite size is at  $\pm 0.1887$ based on quantum Monte Carlo method (QMC)~\cite{braiorrorrs2016phase}.  \label{fig:all-mf-3d}}
\end{figure*}

\begin{figure*}[!htb]
\centerline{\includegraphics[width=\textwidth]{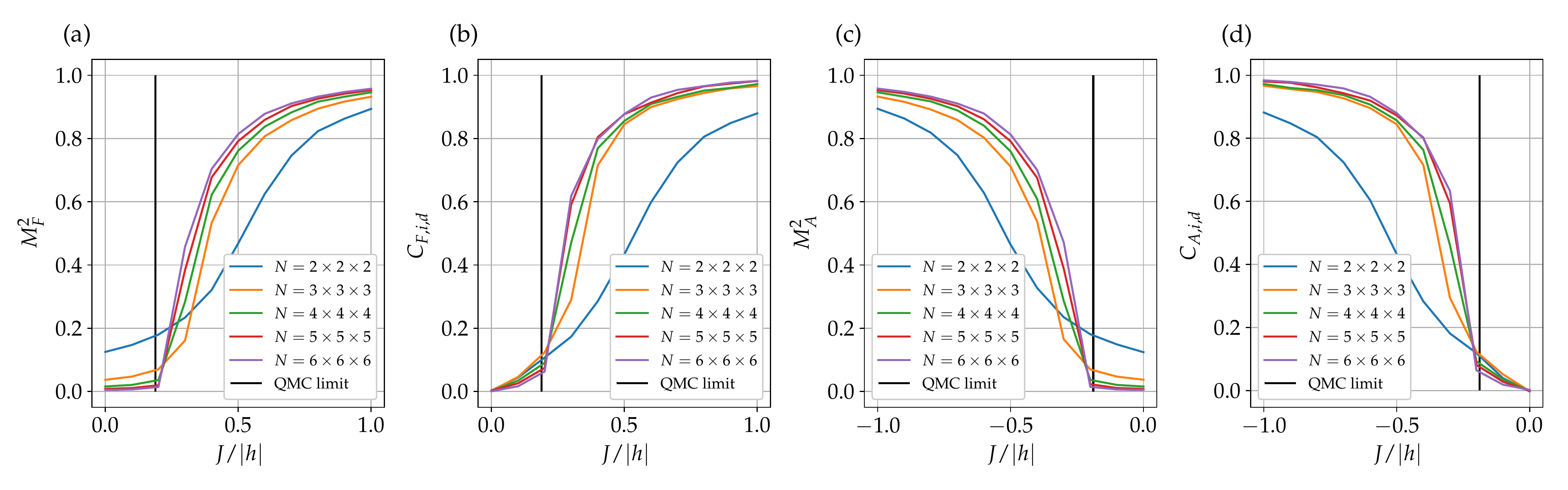}}
\caption{The value of order parameters with RBM-NQS-CS for three-dimensional systems for $J/|h|$ within the range $[-1,0]$ and $[0,1]$, for antiferromagnetic and ferromagnetic order parameters respectively, with $0.1$ intervals and for system with size $N=\{2\times2\times2, 3\times3\times3, 4\times4\times4, 5\times5\times5, 6\times6\times6 \}$. (a), (b), (c) and (d) shows the ferromagnetic magnetisation $M_F^2$, ferromagnetic correlation $C_{F,i,d}$, antiferromagnetic magnetisation $M_A^2$ and antiferromagnetic correlation $C_{A,i,d}$ order parameter, respectively. The critical point at the limit of infinite size is at  $\pm 0.1887$ based on quantum Monte Carlo method (QMC)~\cite{braiorrorrs2016phase}.  \label{fig:all-mf-3d-cs}}
\end{figure*}

Figure~\ref{fig:all-mf-3d} and Figure~\ref{fig:all-mf-3d-cs}  shows the value of the order parameters for a three-dimensional system with RBM-NQS-IT and RBM-NQS-CS, respectively. We calculate the order parameters for $J/|h|$ within the range $[-1,0]$ and $[0,1]$, for antiferromagnetic and ferromagnetic order parameters respectively, with $0.1$ intervals and for systems with sizes $N=\{2\times 2 \times 2,3 \times 3 \times 3,4 \times 4 \times 4,5\times 5\times 5,6\times 6 \times 6\}$.

\section{Effectiveness of finding the inflection point for a system of a given size for correlation order parameters}
\label{sec:size_eval_app_quant}

Table~\ref{tab:qcp_1d_cf} and Table~\ref{tab:qcp_1d_ca} shows the value of the inflection point  for different size of one-dimensional, two-dimensional and three-dimensional systems with RBM-NQS-CS, RBM-NQS-IT and tensor network method with ferromagnetic correlation $C_{F,i,d}$ and  antiferromagnetic correlation $C_{A,i,d}$ order parameter, respectively. 

\section{Effectiveness of finding the  inflection point at the limit of infinite size for correlation order parameters}
\label{sec:limit_eval_app}
\begin{figure}[!htb]
\centerline{\includegraphics[width=0.46\textwidth]{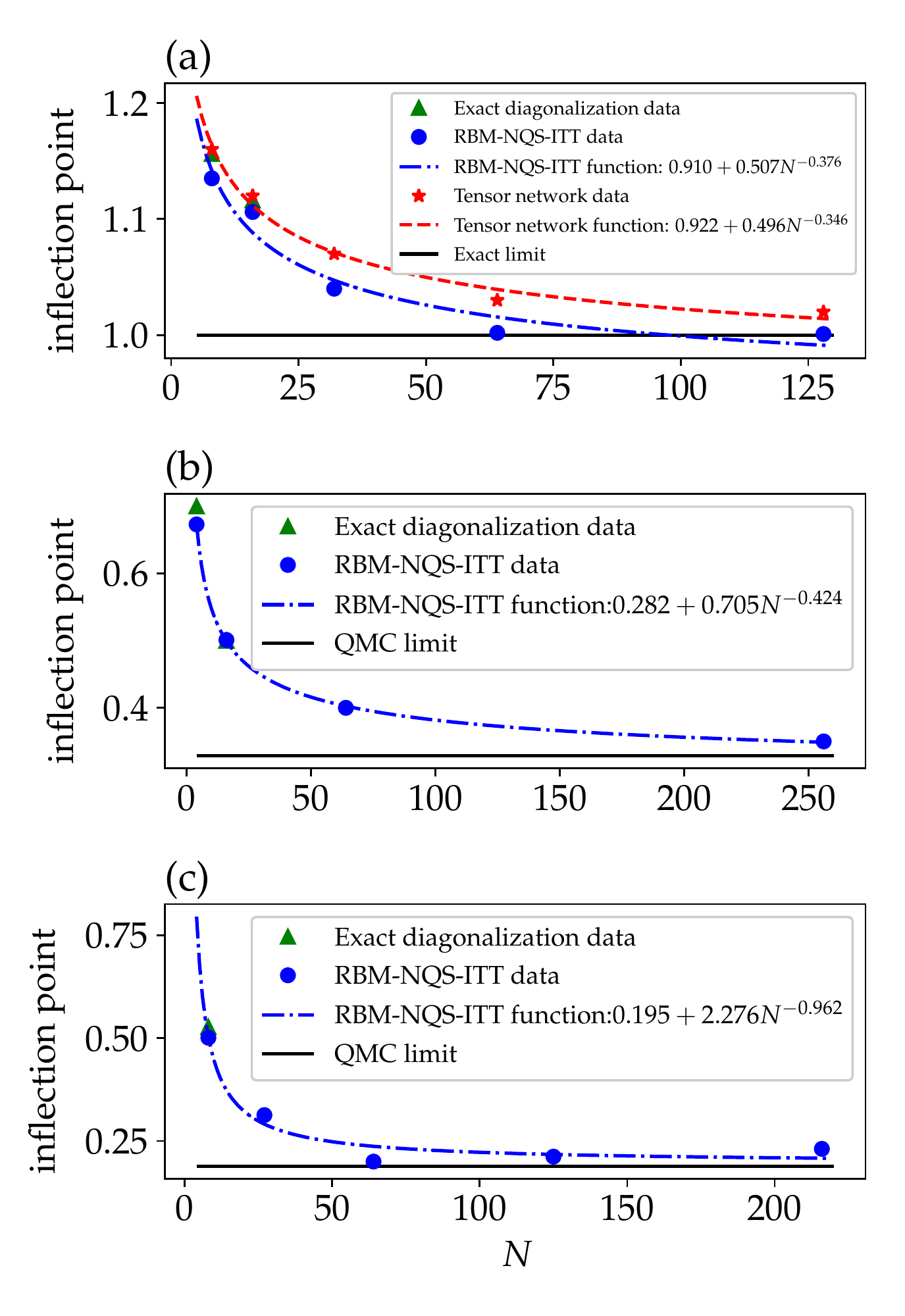}}
\caption{The evaluation of the critical point at the limit of infinite size by fitting the inflection points  as a function of the size of the system in one- (a), two- (b), three- (c) dimensional models.  We use the ferromagnetic correlation $C_{F,i,d}$ to find the critical point. The critical point at the limit of infinite size is at $J/|h|=1$~\cite{suzuki2012quantum} in one-dimensional system, $J/|h|=0.32847$~\cite{blote2002cluster} in two-dimensional system and $J/|h|=0.1887$~\cite{braiorrorrs2016phase} in three-dimensional system.  \label{fig:limit-1d-czf} }
\end{figure}

\begin{figure}[!htb]
\centerline{\includegraphics[width=0.45\textwidth]{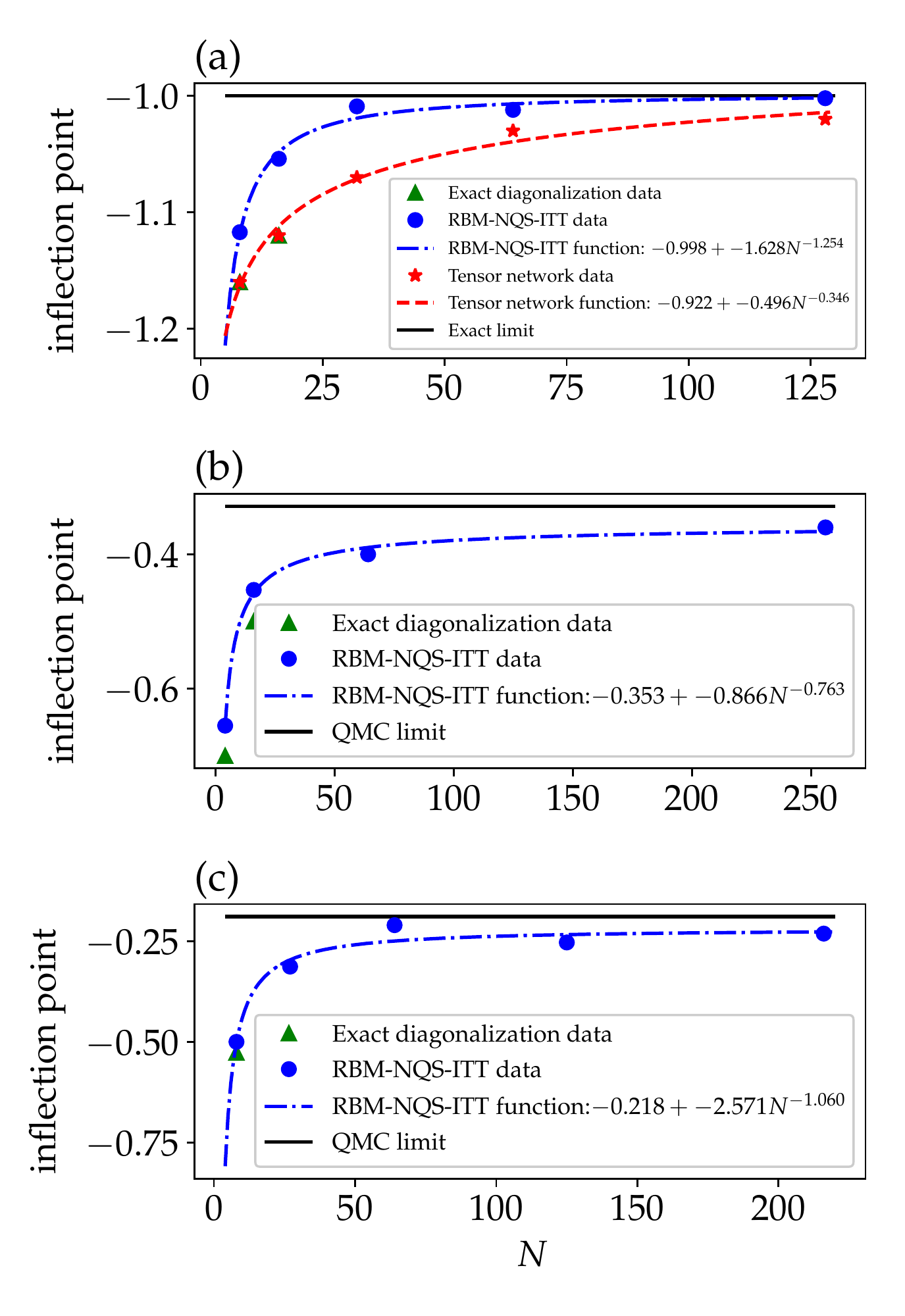}}
\caption{The evaluation of the critical point at the limit of infinite size by fitting the inflection points  as a function of the size of the system in one- (a), two- (b), three- (c) dimensional models.  We use the antiferromagnetic correlation $C_{A,i,d}$ to find the critical point. The critical point at the limit of infinite size is at $J/|h|=-1$~\cite{suzuki2012quantum} in one-dimensional system, $J/|h|=-0.32847$~\cite{blote2002cluster} in two-dimensional system and $J/|h|=-0.1887$~\cite{braiorrorrs2016phase} in three-dimensional system.  \label{fig:limit-1d-cza}} 
\end{figure}

Figure~\ref{fig:limit-1d-czf} (a), (b) and (c) show the evaluation of the critical point at the limit of infinite size by fitting the inflection points  as a function of the size of the system in the one-dimensional, two-dimensional and three-dimensional model, respectively, with ferromagnetic correlation $C_{F,i,d}$ order parameter. Figure~\ref{fig:limit-1d-cza} shows the same evaluation with antiferromagnetic correlation $C_{A,i,d}$ order parameter.

\twocolumn

\begin{table}
\begin{center}
 {\caption{The value of the inflection point  for one-, two- and three-dimensional systems of given sizes with RBM-NQS-CS, RBM-NQS-IT, tensor network and exact diagonalization method with ferromagnetic correlation $C_{F,i,d}$ order parameter. The value inside the parentheses is the standard deviation.}\label{tab:qcp_1d_cf}\label{tab:qcp_2d_cf}\label{tab:qcp_3d_cf}}

\begin{tabular}{ccccc}
\toprule
\textbf{\begin{tabular}[c]{@{}c@{}}System \\ size\end{tabular}} & \textbf{\begin{tabular}[c]{@{}c@{}}RBM-NQS-CS\end{tabular}} & \textbf{\begin{tabular}[c]{@{}c@{}}RBM-NQS-IT\end{tabular}} & \textbf{\begin{tabular}[c]{@{}c@{}}Tensor\\ network\end{tabular}} & \textbf{\begin{tabular}[c]{@{}c@{}}Exact\\ diag.\end{tabular}} \\ \midrule
$8$                                                               & 1.107 (0.07)        & 1.135 (0.005)       & 1.16                                                              & 1.156                                                          \\ \midrule
$16$                                                              & 1.084 (0.04)        & 1.106 (0.007)       & 1.12                                                              & 1.116                                                          \\ \midrule
$32$                                                              & 1.007 (0.05)        & 1.040 (0.012)       & 1.07                                                              & -                                                              \\ \midrule
$64$                                                              & 0.524 (0.32)        & 1.002 (0.009)       & 1.03                                                              & -                                                              \\ \midrule
$128$                                                             & 0.632 (0.43)        & 1.001 (0.005)       & 1.01                                                              & -                                                              \\ \midrule \midrule
$2 \times 2$                                                   & 0.611 (0.102)       & 0.673 (0.05)        & -                                                                 & 0.7                                                            \\ \midrule
$4 \times 4$                                                    & 0.428 (0.041)       & 0.501 (0.003)       & -                                                                 & 0.5                                                            \\ \midrule \midrule
$2 \times 2 \times 2$                                          & 0.501 (0.002)       & 0.502 (0.001)       & -                                                                 & 0.527                                                         
                                                                  
\\ \bottomrule
\end{tabular}         
\end{center}

\end{table}

\begin{table}
\begin{center}
\setlength{\abovecaptionskip}{-6pt}
 {\caption{The value of the inflection point  for one-, two- and three-dimensional systems of given sizes with RBM-NQS-CS, RBM-NQS-IT, tensor network and exact diagonalization method with antiferromagnetic correlation $C_{A,i,d}$ order parameter. The value inside the parentheses is the standard deviation.}\label{tab:qcp_1d_ca}\label{tab:qcp_2d_ca}\label{tab:qcp_3d_ca}}

\begin{tabular}{ccccc}
\toprule
\textbf{\begin{tabular}[c]{@{}c@{}}System \\ size\end{tabular}} & \textbf{\begin{tabular}[c]{@{}c@{}}RBM-NQS-CS\end{tabular}} & \textbf{\begin{tabular}[c]{@{}c@{}}RBM-NQS-IT\end{tabular}} & \textbf{\begin{tabular}[c]{@{}c@{}}Tensor\\ network\end{tabular}} & \textbf{\begin{tabular}[c]{@{}c@{}}Exact\\ diag.\end{tabular}} \\ \midrule
$8$                                                             & -1.074 (0.09)       & -1.117 (0.007)      & -1.16                                                             & -1.109                                                         \\ \midrule
$16$                                                            & -1.103 (0.03)       & -1.054 (0.006)      & -1.12                                                             & -1.090                                                         \\ \midrule
$32$                                                            & -1.016 (0.007)      & -1.009 (0.009)      & -1.07                                                             & -                                                              \\ \midrule
$64$                                                            & -0.910 (0.23)       & -1.012 (0.009)      & -1.03                                                             & -                                                              \\ \midrule
128                                                             & -0.413 (0.27)       & -1.002 (0.002)      & -1.02                                                             & -                                                              \\ \midrule\midrule
$2 \times 2$                                                    & -0.617 (0.07)       & -0.655 (0.05)       & -                                                                 & -0.7                                                           \\ \midrule
$4 \times 4$                                                    & -0.424 (0.03)       & -0.453 (0.05)       & -                                                                 & -0.5                                                           \\ \midrule\midrule
$2 \times 2 \times 2$                                           & -0.501 (0.003)      & -0.502 (0.002)      & -                                                                 & -0.527

\\ \bottomrule
\end{tabular}         
\end{center}

\end{table}

\end{document}